\documentclass[3p]{elsarticle}
\usepackage{lineno,hyperref}
\usepackage{graphicx}
\usepackage{amsmath}
\usepackage{color}
\usepackage[normalem]{ulem}
\usepackage{ mathrsfs }
\journal{arXiv}
\begin{document}

\begin{frontmatter}

\title{Distributions of Historic Market Data -- Relaxation and Correlations}

\author[mymainaddress]{M. Dashti Moghaddam}
\author[mymainaddress]{Zhiyuan Liu}
\author[mymainaddress]{R. A. Serota\fnref{myfootnote}}
\fntext[myfootnote]{serota@ucmail.uc.edu}

\address[mymainaddress]{Department of Physics, University of Cincinnati, Cincinnati, Ohio 45221-0011}

\begin{abstract}
We investigate relaxation and correlations in a class of mean-reverting models for stochastic variances. We derive closed-form expressions for the correlation functions and leverage for a general form of the stochastic term. We also discuss correlation functions and leverage for three specific models -- multiplicative, Heston (Cox-Ingersoll-Ross) and combined multiplicative-Heston -- whose steady-state probability density functions are Gamma, Inverse Gamma and Beta Prime respectively, the latter two exhibiting "fat" tails. For the Heston model, we apply the eigenvalue analysis of the Fokker-Planck equation to derive the correlation function -- in agreement with the general analysis -- and to identify a series of time scales, which are observable in relaxation of cumulants on approach to the steady state. We test our findings on a very large set of historic financial markets data. 
\end{abstract}


\end{frontmatter}

\section{Introduction\label{Introduction}}

Questions about correlations between and relaxation of quantities described by stochastic differential equations (SDE) have a very long history in physics applications \cite{uhlenbeck1930theory}, \cite{schenzle1979multiplicative}. More recently, they found a new urgency in areas related to economics   \cite{bouchaud2000wealth,bouchaud2015growth,ma2013distribution,liu2017correlation,liu2018absence} and finance \cite{perello2002correlated,perello2003random,perello2004multiple,ma2014model}, some of which utilized models originally found in physics. In the most general formulation, one is interested in correlations in the time series generated by a stochastic process, described by an SDE, and in the time scales for relaxation to its steady state. Ideally, one would obtain an analytical expression for the correlation function in terms of the parameters of an SDE and would identify quantities that analytically describe relaxation.

A common purpose of an SDE is to model empirical time series, such as stock prices. Stochastic models for stock returns use stochastic volatility as one of their inputs. In this paper we concentrate on a class of models for stochastic variance -- squared stochastic volatility -- which are characterized by the Generalized Beta Prime steady-state probability density function and its limits corresponding to their mean-reverting subset: Inverse Gamma, Gamma and Beta Prime distributions. We use readily-available historic stock prices data to test our predictions with respect to the correlation functions and leverage, which we derive analytically for these models. In particular, we study both daily and multi-day correlations and leverage.

This paper is organized as follows. In Section \ref{CorrLev}, we identify equations for the covariance of stochastic variance and for the leverage for a general form of stochastic term. We show that the correlation function of stochastic variance depends only on the relaxation parameter. We relate correlations of realized variance for daily and multi-day returns to correlations of stochastic variance. In Section \ref{Models}, we proceed to apply general equations obtained in Section \ref{CorrLev} to specific stochastic terms of mean-reverting models -- multiplicative, Heston, and combined multiplicative-Heston -- and derive their parameters from the historic market data. In \ref{corrdxt2}, we continue the discussion of correlations of multi-day returns introduced in Section \ref{CorrLev}. In \ref{CIRmodel}, we discuss Heston model in greater detail: we find correlations of stochastic variance using eigenvalues analysis of the Fokker-Planck equation as well as study the relaxation of cumulants and the distribution of relaxation times.

\section{Correlations of Stochastic Variance and Leverage \label{CorrLev}}

Equation for de-trended stock log returns can be written as \cite{dashti2018combined}
\begin{equation}
\mathrm{d}x_{t} = \sigma_t \mathrm{d}W_{t}^{(1)} 
\label{xt}
\end{equation}

where $\mathrm{d}W_t$ is a normally distributed Wiener process and $\sigma_t$ is the stochastic volatility which is related to the stochastic variance $v_t$ by $v_t=\sigma_t^{2}$. A general mean-reverting model for the stochastic variance can be written as 

\begin{equation}
\mathrm{d}v_t = -\gamma(v_t - \theta)\mathrm{d}t + g(v_t)\mathrm{d}W_t^{(2)}
\label{general}
\end{equation}
and rewritten as
\begin{equation}
v_t = \theta +\int_{-\infty}^{t}e^{-\gamma(t-t^\prime)} g(v_t)\mathrm{d}W_t^{(2)}
\label{vsolve}
\end{equation}
It is assumed that $\mathrm{d}W_t^{(1)}$ and $\mathrm{d}W_t^{(2)}$ are cross-correlated, with the coefficient $\rho$, as
\begin{equation}
\mathrm{d}W_t^{(2)} = \rho \mathrm{d}W_t^{(1)} + \sqrt{1-\rho^2} \mathrm{d}Z_t
\label{corr}
\end{equation}
where $\mathrm{d}Z_t$ is independent of $\mathrm{d}W_t^{(1)}$. In (\ref{general}), $\gamma$ is the relaxation parameter: $\gamma^{-1}$ is the time scale for achieving the steady-state distribution of $v$ \cite{liu2017correlation}, whose mean value is $\theta$,
\begin{equation}
<v_t>=\theta
\label{meanv}
\end{equation}
From (\ref{xt})) and \ref{meanv} we also have
\begin{equation}
<dx_t^{2}> = <v_t>\mathrm{d}t = \theta \mathrm{d}t
\label{expxt2}
\end{equation}
which directly relates $\theta$ to stock returns data. 

\subsection{Correlation Function of Stochastic Variance \label{Variance}}

Using (\ref{vsolve}), we find the covariance of stochastic variance as
\begin{equation}
cov[v_t v_{t+\tau}]=<v_t v_{t+\tau}>-<v_t>^2=var[v_t] e^{-\gamma \tau}
\label{meanVarCor}
\end{equation}
where 
\begin{equation}
var[v_t]=<v_t^2>-<v_t>^2=\frac{<g^{2}(v_t)>}{2\gamma}
\label{varvt}
\end{equation}
so that the correlation function (Pearson correlation coefficient) depends only on the relaxation parameter
\begin{equation}
corr[v_t v_{t+\tau}]=\frac{ <v_t v_{t+\tau}>-<v_t>^2}{var[v_t]} =e^{-\gamma \tau}
\label{corrvt}
\end{equation}
To obtain $corr[v_t v_{t+\tau}]$ from stock returns we observe that from (\ref{xt}) 
\begin{equation}
<\mathrm{d}x_t^2 \mathrm{d}x_{t+\tau}^2>  = < \sigma_t\mathrm{d}W_t^{(1)} \sigma_t\mathrm{d}W_t^{(1)}  \sigma_{t+\tau} \mathrm{d}W_{t+\tau}^{(1)}  \sigma_{t+\tau}\mathrm{d}W_{t+\tau}^{(1)}>
\label{dxt2corrW}
\end{equation}
which yields 
\begin{equation}
<dx_t^2 dx_{t+\tau}^2> =
\begin{cases}
<v_t v_{t+\tau}>\left( t^2 + 2(t-\tau)^2 \right) & \text{ $t 	\geq \tau$} \\
<v_t v_{t+\tau}>t^2  & \text{ $t \leq  \tau$} 
\end{cases}
\label{dxt2corrmulti}
\end{equation}
for $\tau > 0$ and 
\begin{equation}
<\mathrm{d}x_t^4> = 3 <v_t ^2> t^2
\label{dxt4}
\end{equation}
for $\tau = 0$. The factor of 3 in (\ref{dxt4}) is purely combinatorial and is model-independent. (In general, $<\mathrm{d}x_t^{2n}> = (2n-1)!! <v_t ^n> \mathrm{d}t^n$ \cite{dashti2018combined}). In (\ref{dxt2corrmulti}) and (\ref{dxt4}) we replaced $\mathrm{d}t$ with $t$ -- the number of days accumulation of returns. In what follows we will use $\mathrm{d}t$ and $t$ interchangebly. Specifically for daily returns, $\mathrm{d}t=t=1$, the second equation in (\ref{dxt2corrmulti}) is the one obtained in \cite{perello2004multiple}. 
It follows then from (\ref{expxt2}) and (\ref{corrvt})-(\ref{dxt4}) that for daily returns
\begin{equation}
\frac{<\mathrm{d}x_t^2 \mathrm{d}x_{t+\tau}^2 >- <\mathrm{d}x_t^2>^2}{\frac{1}{3} <\mathrm{d}x_t^4> - <\mathrm{d}x_t^2>^2}=e^{-\gamma \tau}
\label{daily}
\end{equation}

\begin{figure}[!htbp]
\centering
\begin{tabular}{cc}
\includegraphics[width = 0.49 \textwidth]{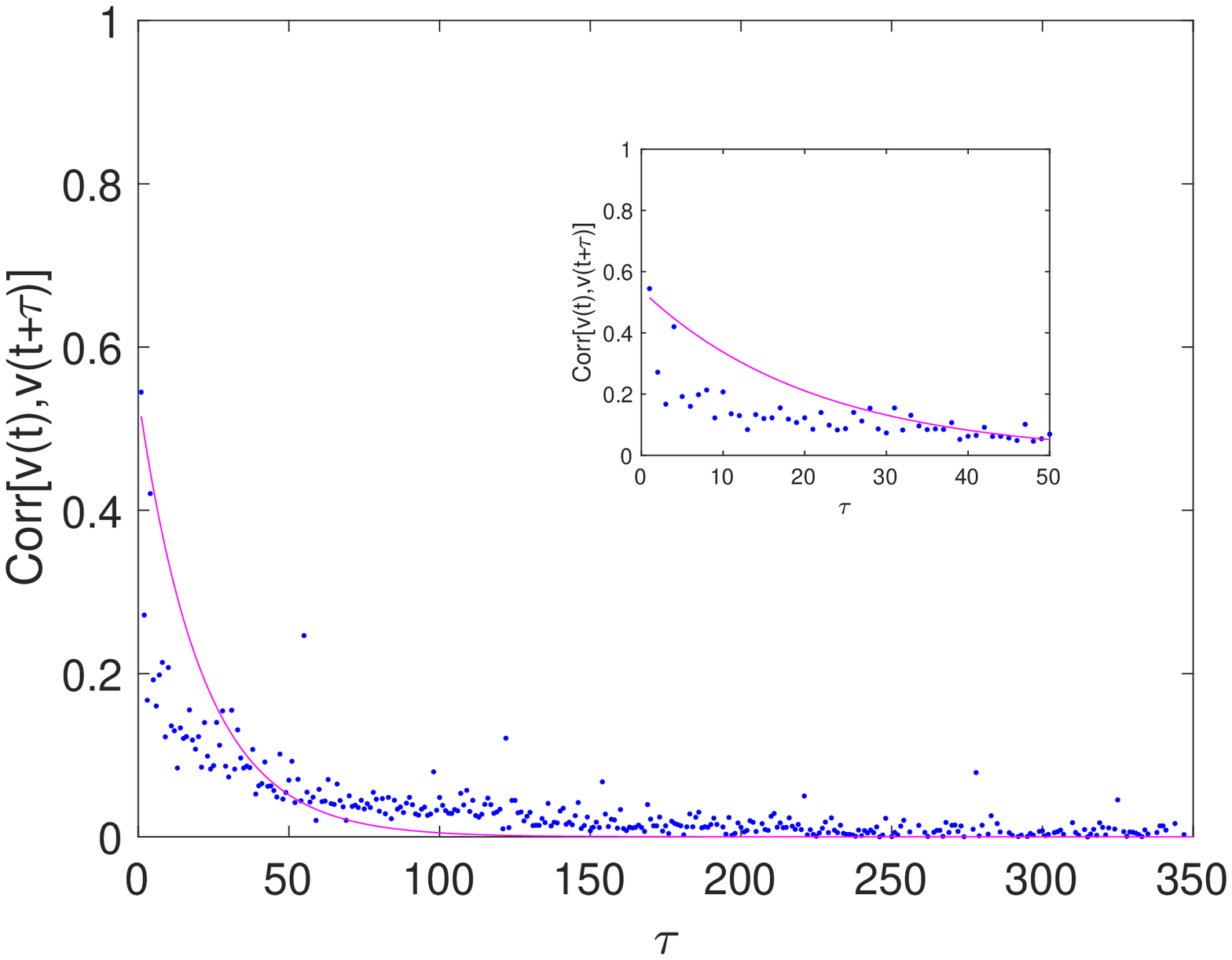}
\includegraphics[width = 0.49 \textwidth]{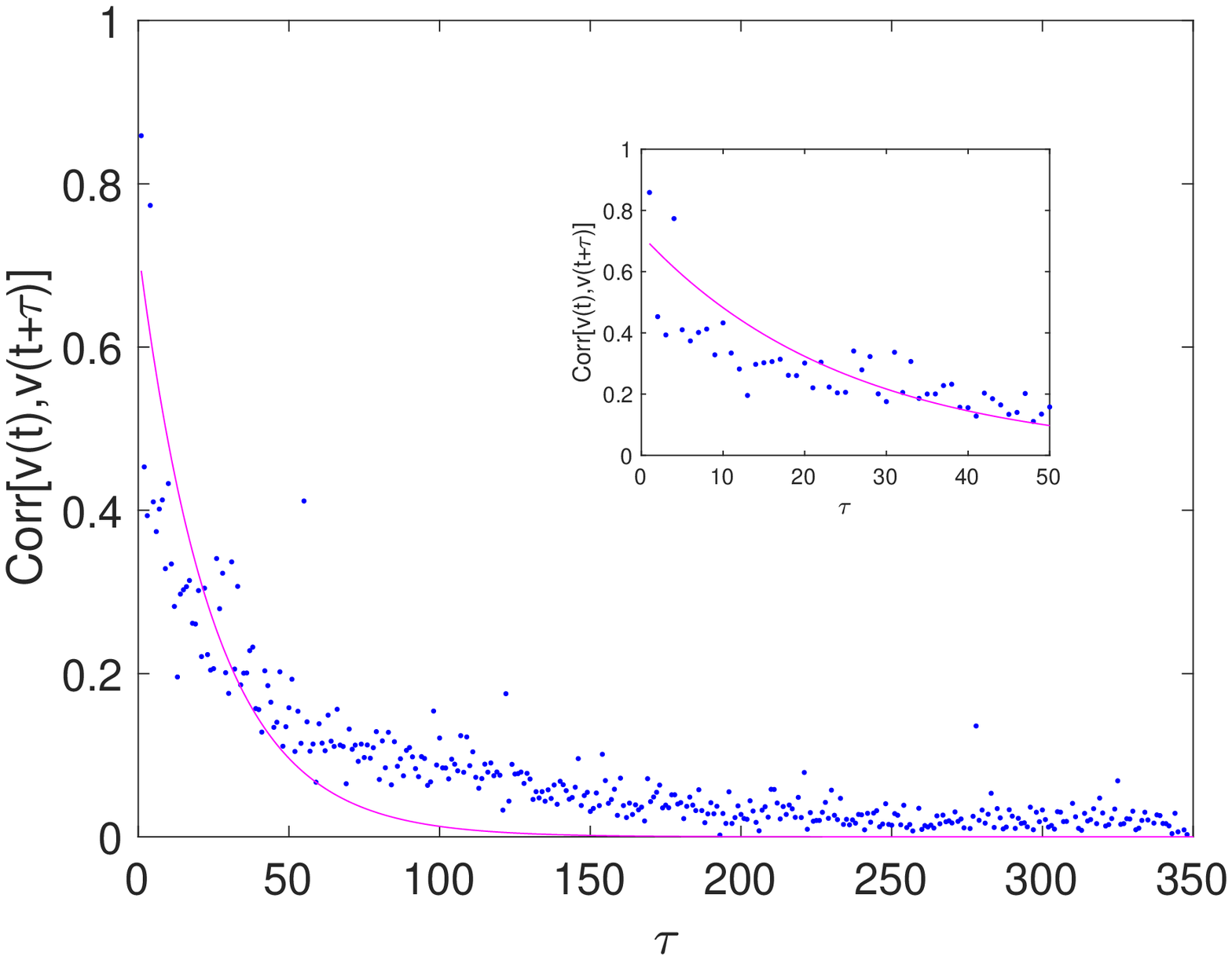}
\end{tabular}
\caption{Correlation function of stochastic variance per l.h.s. of (\ref{daily}) for daily returns, $\mathrm{d}t=1$, fitted with $a \times exp(-\gamma \tau)$. Left: DJIA, $a = 0.5481,  \gamma = 0.04521$. Right: S\&P500, $a = 0.7219,  \gamma = 0.04031$.}
\label{Correlation}
\end{figure}

\begin{figure}[!htbp]
\centering
\begin{tabular}{cc}
\includegraphics[width = 0.40 \textwidth]{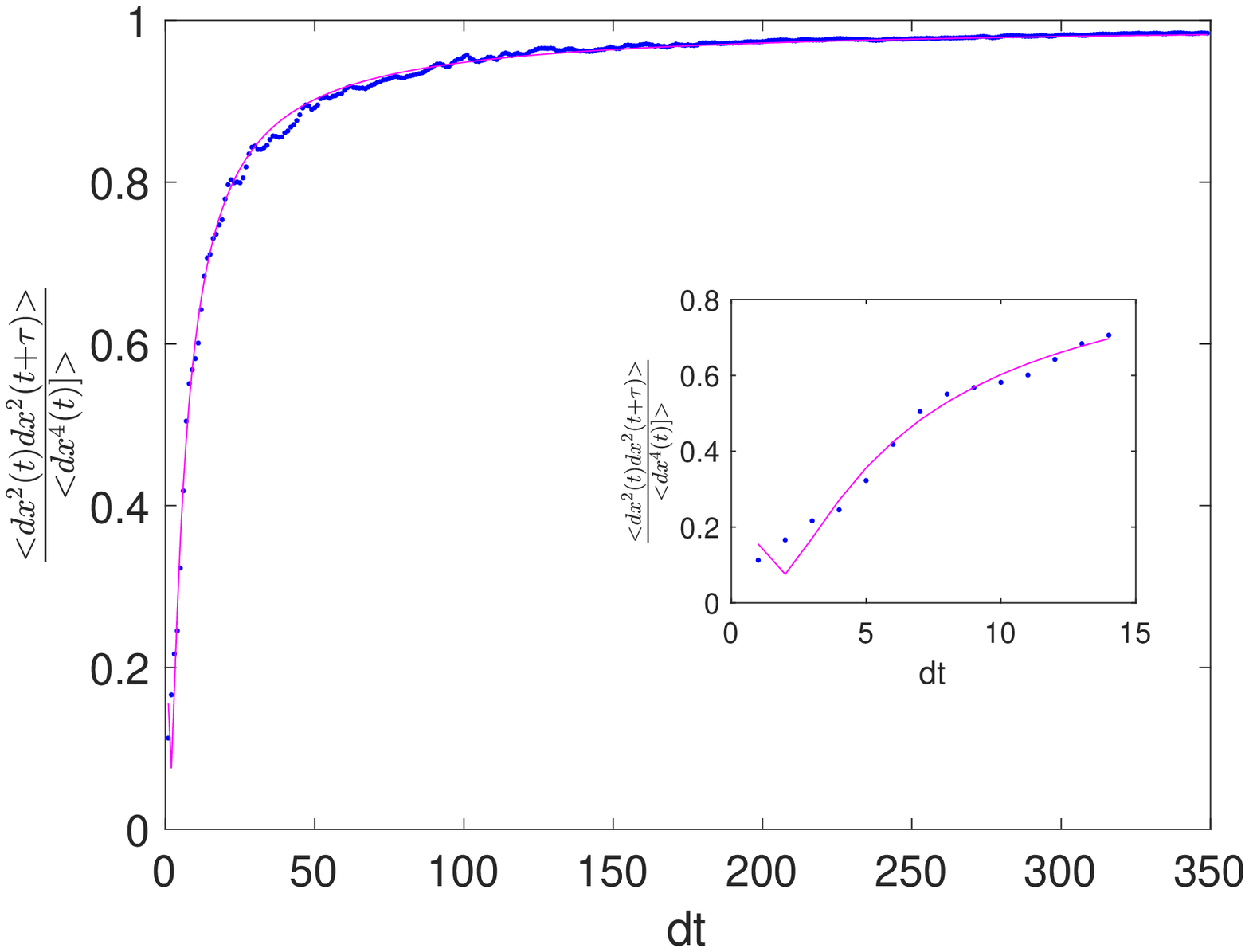}
\includegraphics[width = 0.40 \textwidth]{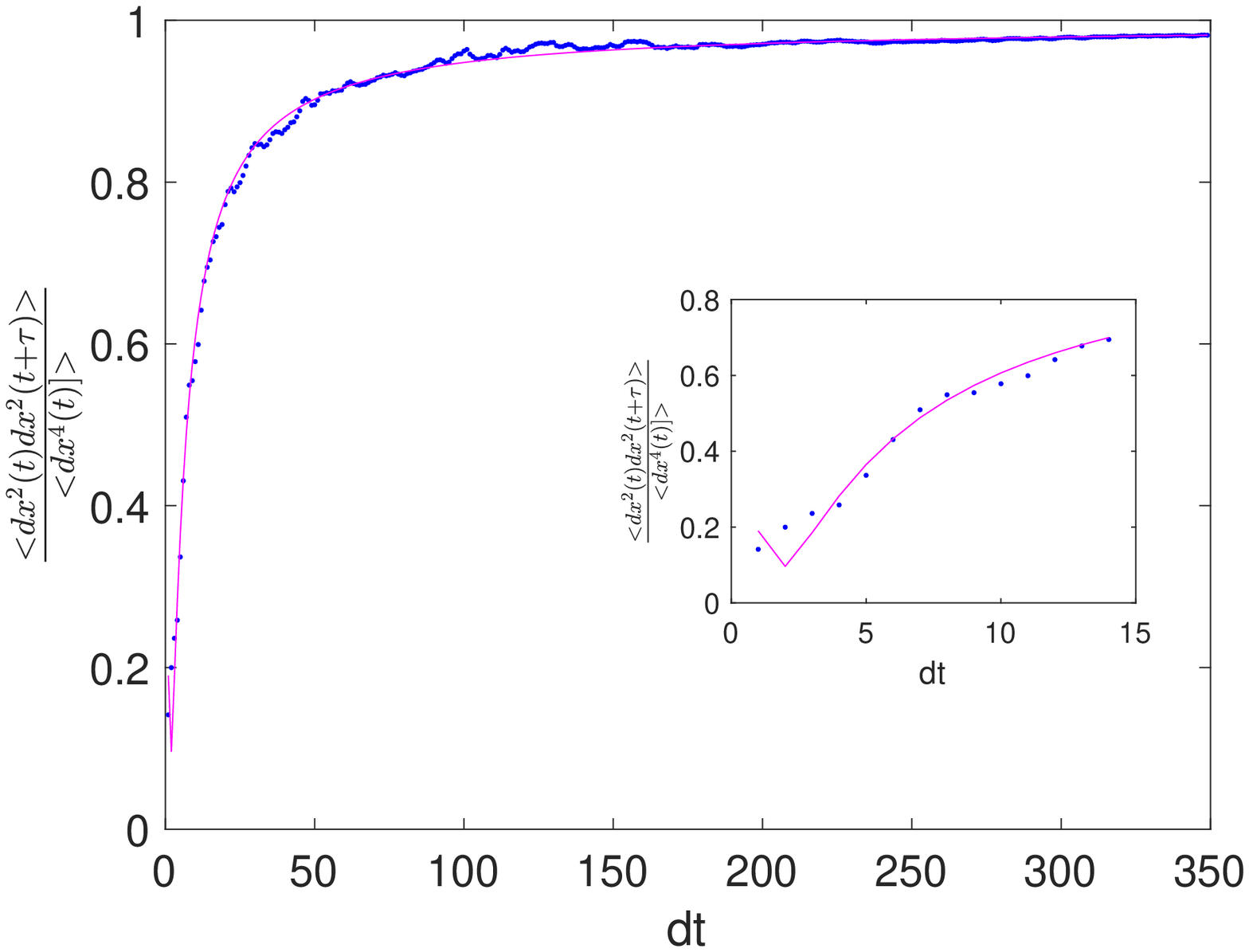} \\ 
\includegraphics[width = 0.40 \textwidth]{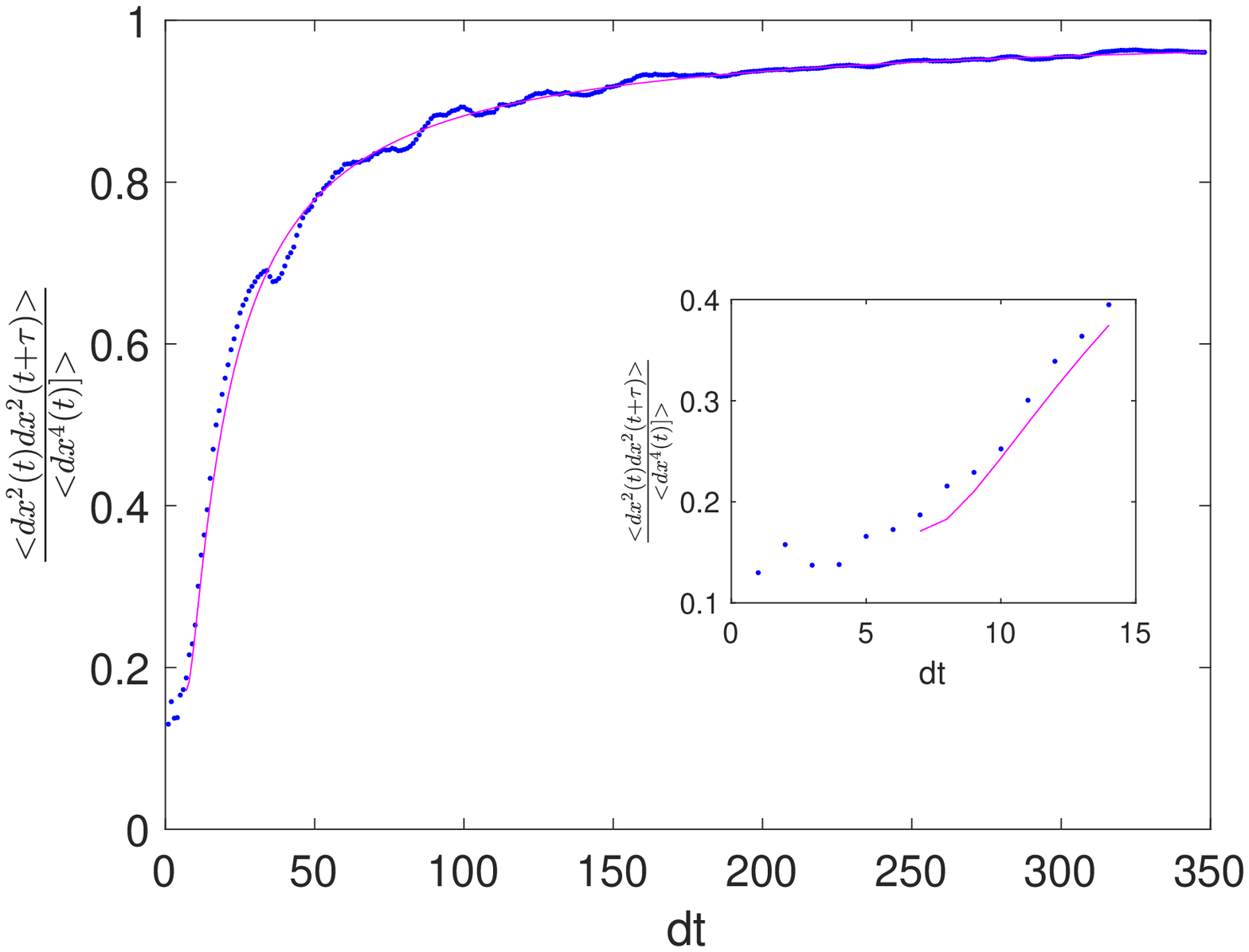}
\includegraphics[width = 0.40 \textwidth]{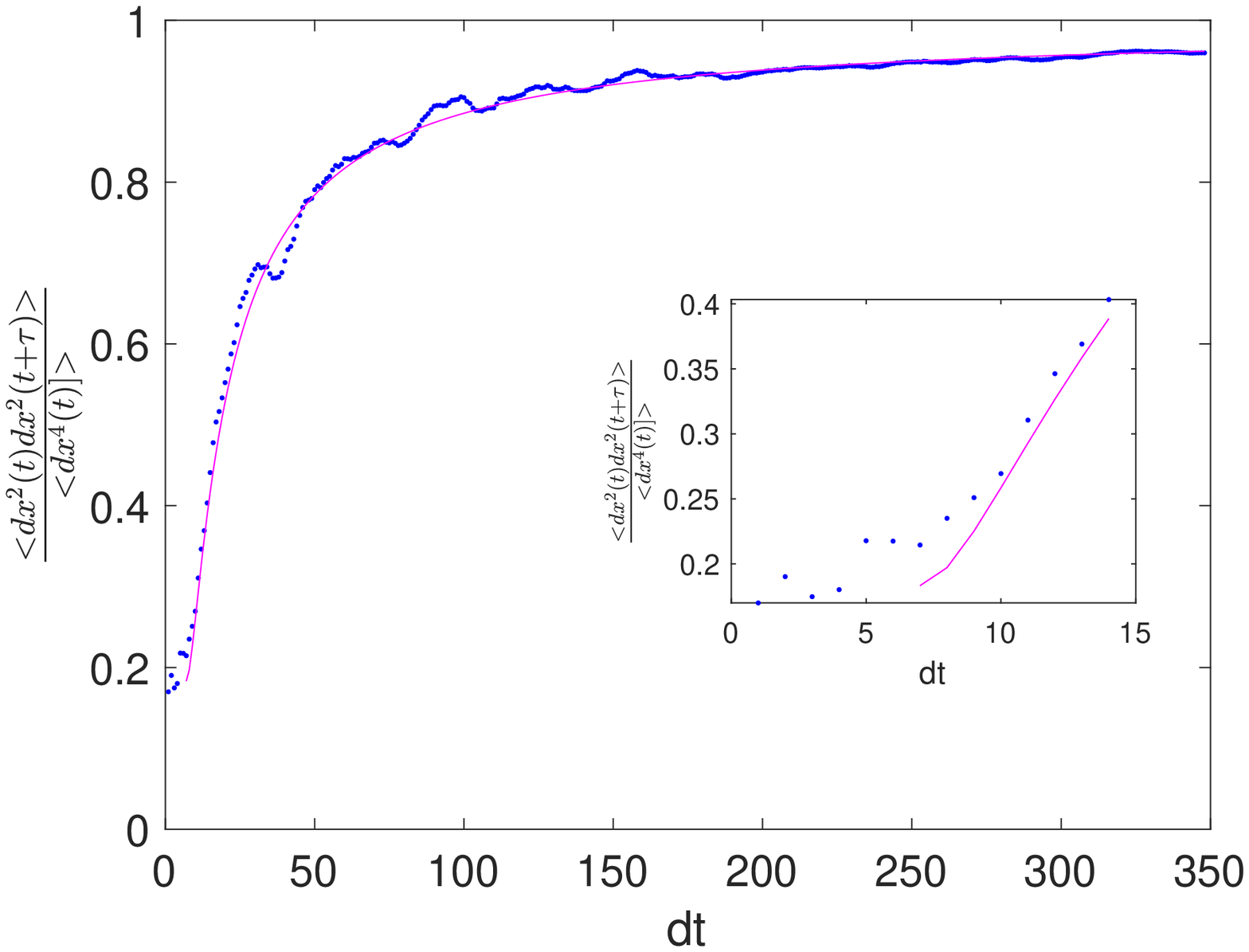} \\
\includegraphics[width = 0.40 \textwidth]{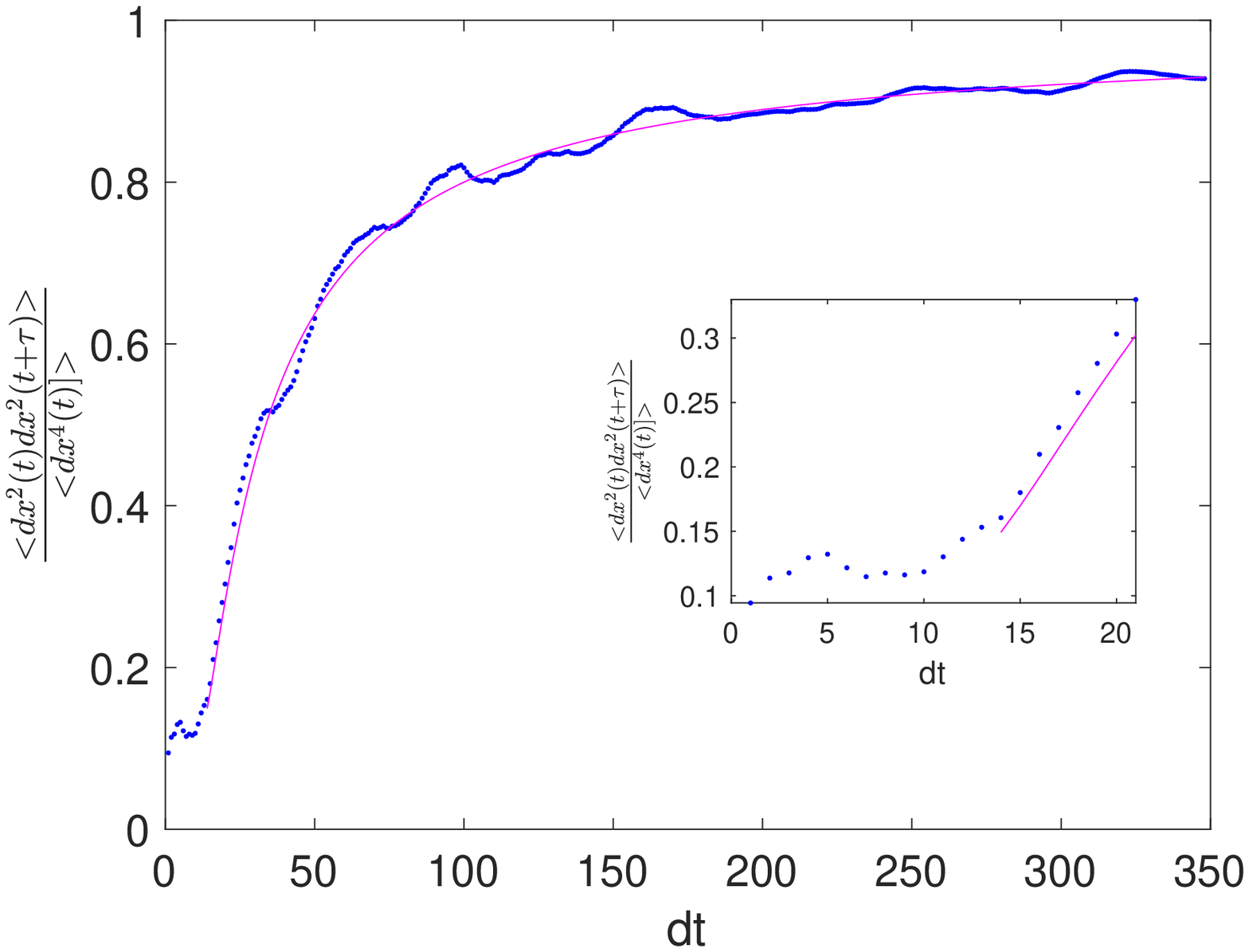}
\includegraphics[width = 0.40 \textwidth]{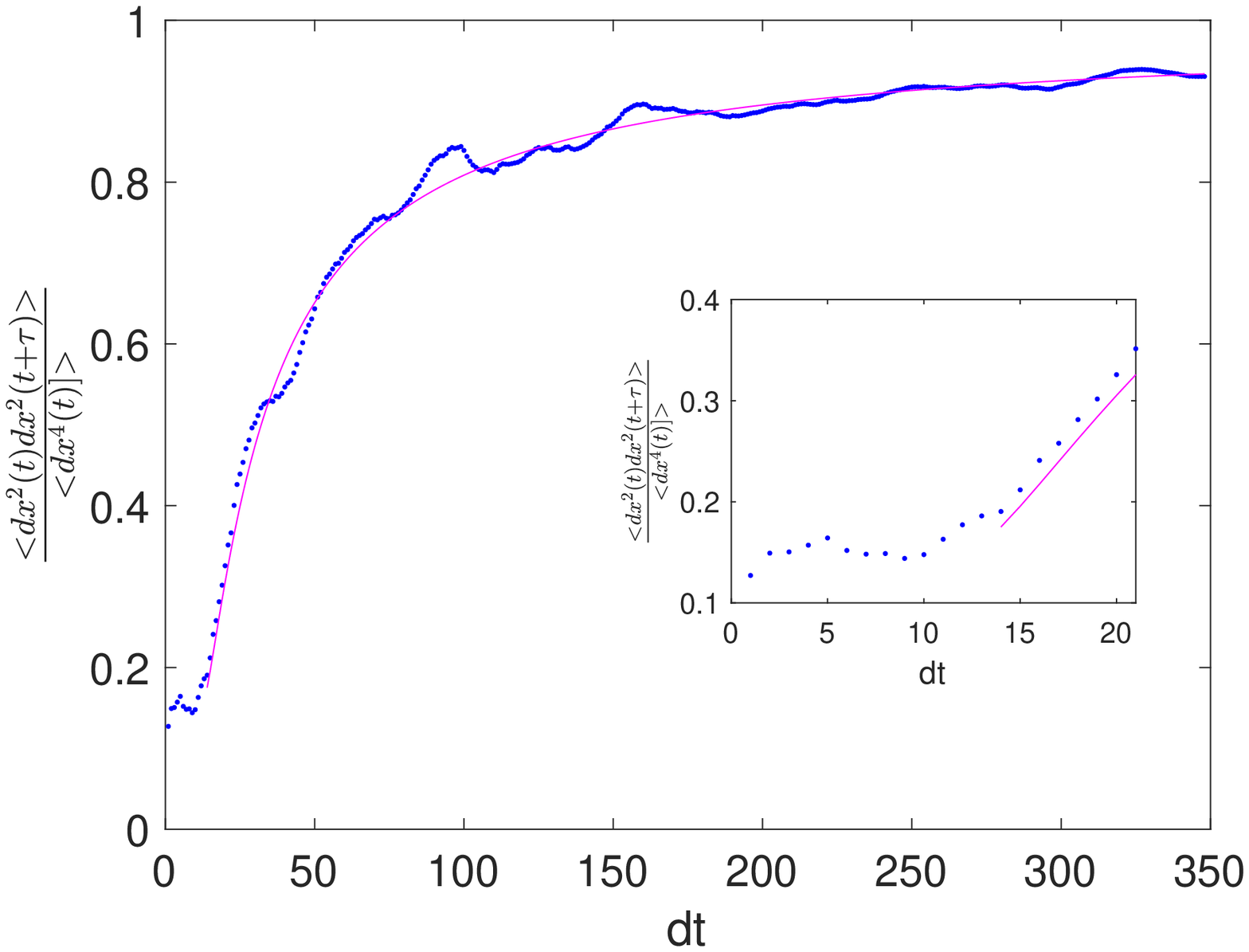} \\
\includegraphics[width = 0.40 \textwidth]{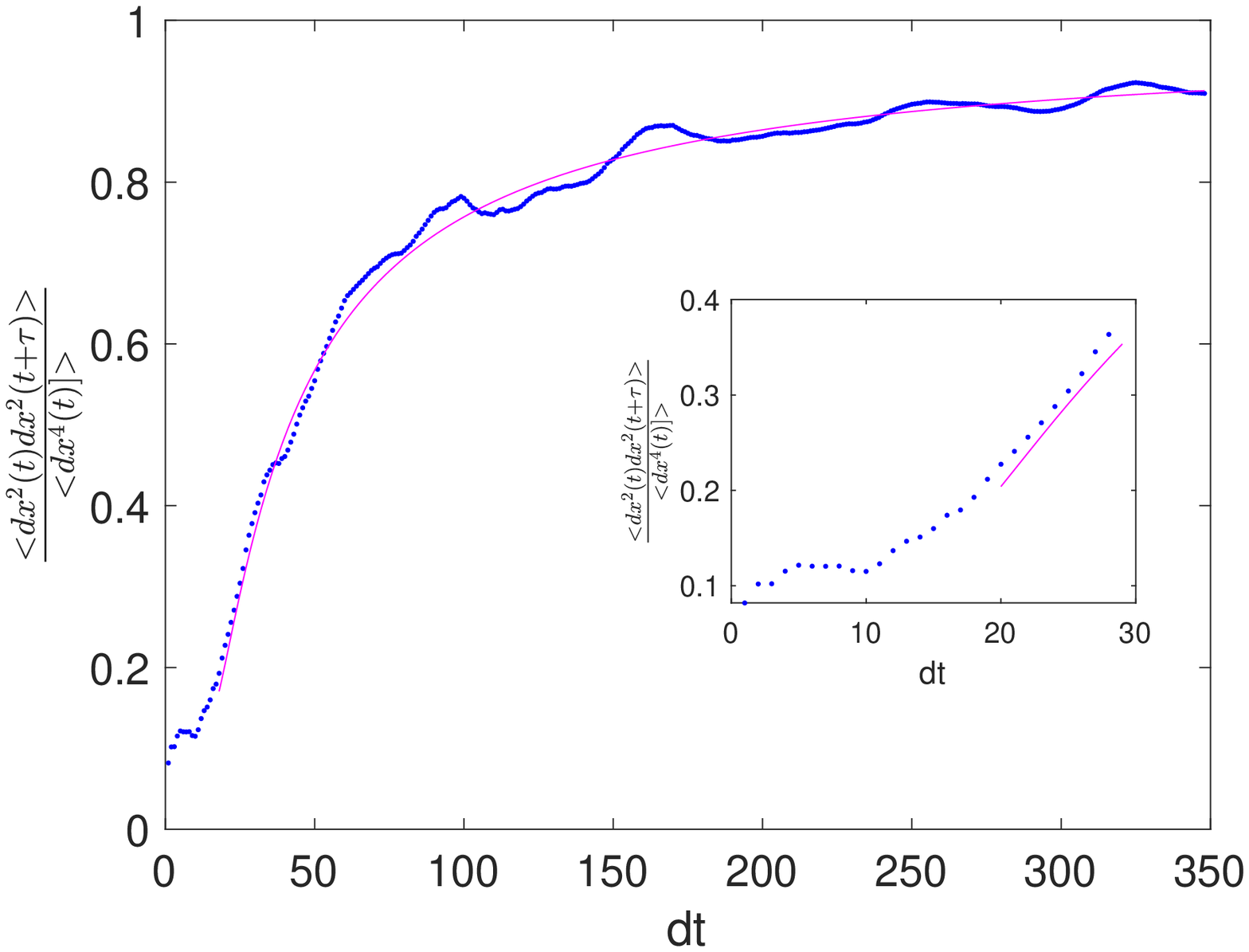}
\includegraphics[width = 0.40 \textwidth]{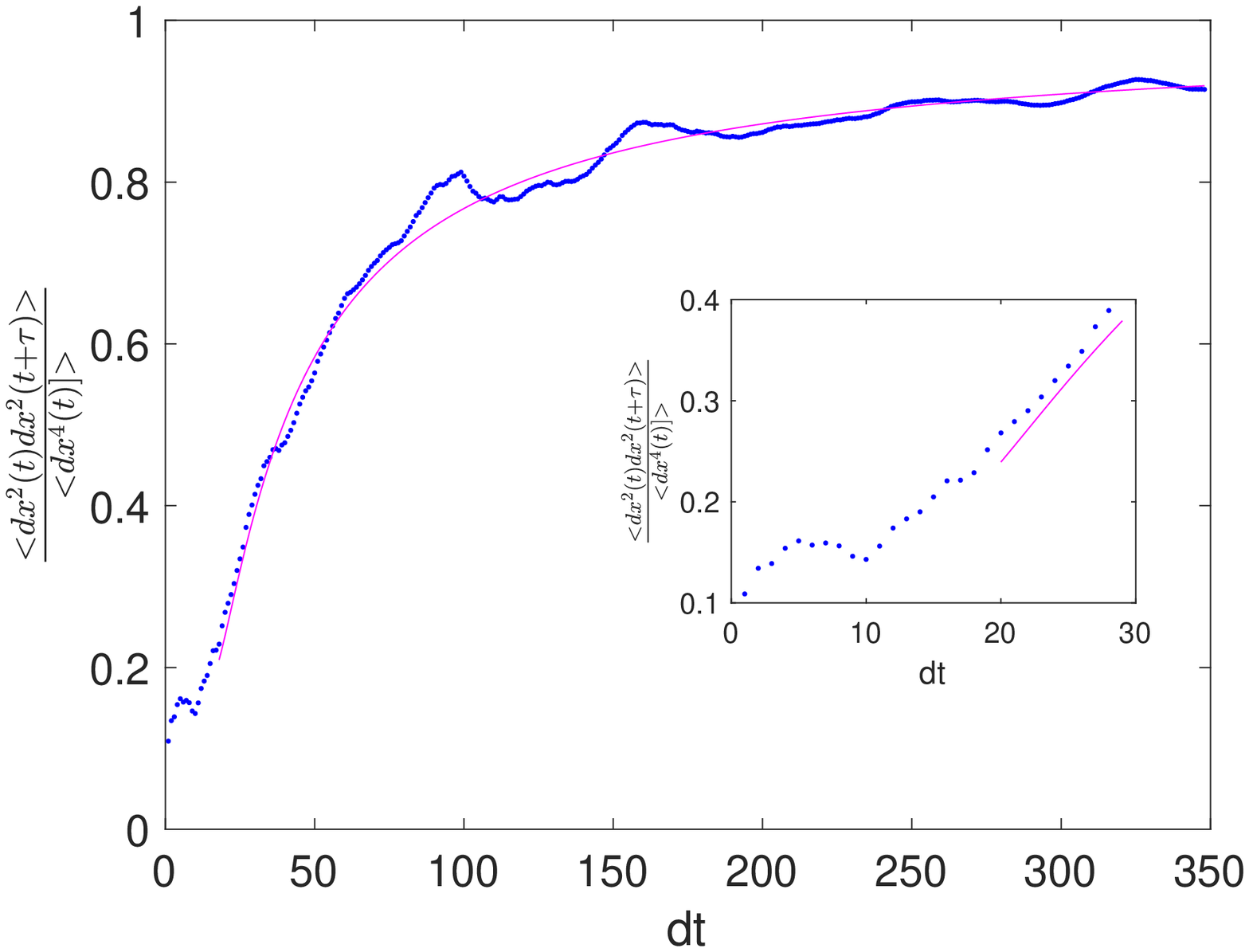} \\

\end{tabular}
\caption{$\frac{<\mathrm{d}x_t^2 \mathrm{d}x_{t+\tau}^2 >}{<\mathrm{d}x_t^4>}$ as a function of $t$. From top row to bottom $\tau=1, 7, 14, 21$. Left column DJIA, right S\&P.}
\label{tau}
\end{figure}

Fig. \ref{Correlation} show plots and their fits for the l.h.s. of (\ref{daily}) for daily returns. It is obvious that the fit is rather poor relative to the analytical prediction. This is mostly likely because mean-reverting, continuous stochastic volatility models are not appropriate for daily returns. On the other hand, such models are more relevant to multi-day returns. Consequently, it is of interest to investigate correlations of multi-day returns. Toward this end, we first discuss the consequences of (\ref{dxt2corrmulti}). From the latter, we find that for $\tau \ll \gamma^{-1} \approx 21$, that is $ <v_t v_{t+\tau}> \approx <v_t^2>$,
\begin{equation}
\frac{<\mathrm{d}x_t^2 \mathrm{d}x_{t+\tau}^2 >}{<\mathrm{d}x_t^4>} \approx
\begin{cases}
1 - \frac{4 \tau}{3t}+  \frac{2 \tau^2}{3t^2} & \text{ $t 	\geq \tau$} \\
\frac{1}{3} & \text{ $t \leq  \tau$} 
\end{cases}
\label{ratio}
\end{equation}
Fig. \ref{tau} shows the dependence of the l.h.s. of (\ref{ratio}) as a function of the number of days of accumulation for $\tau= 1, 7, 14, 21$ respectively and their fits for $t > \tau$ with $a - b \frac{\tau}{t} + c \frac{\tau^2}{t^2} $ with the values of fitting parameters and $r^2$ statistics collected in Table \ref{table}. Clearly, predictions of (\ref{ratio}) hold up quite well. 

\begin{table}[!htb]
\caption{Parameters of $t > \tau$ with $a - b \frac{\tau}{t} + c \frac{\tau^2}{t^2}$ fit for $t \ge \tau$ in Fig. \ref{tau} and their $r^2$ statistics.}
\label{table}
\begin{minipage}{0.5\textwidth}
\begin{center}
\begin{tabular}{c c c c c c} 
\multicolumn{5}{c}{DJIA} \\
\hline
         $ \tau$ &   a &       b &          c& $r^2$   \\
         1   &   0.99  &   4.92  &  6.48   &   0.99\\
         7   &   0.99   &   1.65   &  0.83  &  0.99\\
         14  &   0.98   &  1.40   &  0.56  &  0.99\\
         21 & 0.98 & 1.34 &  0.53  &  0.99\\
\hline
\hline
\end{tabular}
\end{center}
\end{minipage}
\begin{minipage}{.5\textwidth}
\begin{center}
\begin{tabular}{c c c c c c} 
\multicolumn{5}{c}{S\&P500} \\
\hline
          $ \tau$ &a &       b &          c& $r^2$   \\
         1   &   0.99  &   4.87  &  6.52   &   0.99\\
         7   &   0.99   &   1.61   &  0.79  &  0.99\\
         14  &   0.98   &  1.35   &  0.54  &  0.99\\
         21 & 0.98 & 1.30 &  0.53  &  0.99\\
\hline
\hline
\end{tabular}
\end{center}
  \end{minipage}

\end{table}
Due to the complicated nature of relationship between $<dx_t^2 dx_{t+\tau}^2>$ and $<v_t v_{t+\tau}>$ in (\ref{dxt2corrmulti}), it is unclear how to express the result for $corr[v_t v_{t+\tau}]$ in (\ref{corrvt}) in terms of market quantities for multi-day returns. Instead we surmised that $corr[\mathrm{d}x_t^2 \mathrm{d}x_{t+\tau}^2]$ may have a similarly clean dependence on time
\begin{equation}
corr[\mathrm{d}x_t^2 \mathrm{d}x_{t+\tau}^2]=\frac{<\mathrm{d}x_t^2 \mathrm{d}x_{t+\tau}^2> - <\mathrm{d}x_t^2>^2}{<\mathrm{d}x_t^4> -<\mathrm{d}x_t^2>^2} \approx e^{-\gamma_1 \tau}
\label{multiday}
\end{equation}
In  \ref{corrdxt2} we empirically investigate the dependence of $\gamma_1$ on the number of days of return.

\subsection{Leverage \label{Leverage}}
We now turn to leverage effect, whose main "prize" is the cross-correlation $\rho$, but which also allows to independently evaluate $\gamma$. Leverage is defined as 
\begin{equation}
\mathscr{L}(\tau)=\frac{<\mathrm{d}x_{t+\tau}^2 \mathrm{d}x_t>}{<\mathrm{d}x_t^2>^2}
\label{LeverageDef}
\end{equation}
A priori, it is clear that $\rho$ should be negative as upward fluctuations of volatility should lead to downward fluctuations in returns and that it should decay exponentially in time. Market leverage was studied in great detail in \cite{perello2002correlated,perello2003random,perello2004multiple}. We believe that functional derivative in (7) of \cite{perello2003random} can be greatly simplified -- to $exp (-\gamma \tau) g(v_t)$ in our notations -- so that the (\ref{LeverageDef}) reduces to 
\begin{equation}
\mathscr{L}(\tau)=   \frac{\rho <v_t^{1/2} g(v_t)> exp (-\gamma \tau)}{\theta^2}
\label{LeverageSimple}
\end{equation}

\section{Multiplicative, Heston and Combined Models of Stochastic Variance \label{Models}}
\subsection{Analytical results \label{Analytical}}
Expressions \ref{varvt} and \ref{LeverageSimple} in Section \ref{CorrLev} did not specify the form of $g(v_t)$ and it is a priori clear that relaxation of the covariances $\propto exp(-\gamma \tau)$ should depend only on the single relaxation time parameter in the model, $\gamma$. A very general model of stochastic volatility is given by
\begin{equation}
\mathrm{d}v_t = -\gamma(v_t - \theta v_t^{1-\alpha})\mathrm{d}t + \sqrt{\kappa_2^2 v_t^2 + \kappa_\alpha^2 v_t^ {2-\alpha}}\mathrm{d}W_t^{(2)}
\label{GB2SDE}
\end{equation}
Its steady-state distribution (probability density function -- PDF) is a Generalized Beta Prime, or GB2, distribution given by \cite{hertzler2003classical, dashti2018realized, dashti2019generalized, dashti2019modeling} 
\begin{equation}
GB2(v_t; p,q,\beta,\alpha)=\frac{\alpha (1+({\frac{v_t}{\beta}})^{\alpha})^{-p-q}(\frac{v_t}{\beta})^{-1+p\alpha}}{\beta B(p,q)}
\label{GB2x}
\end{equation}
where $B(p,q)$ is a beta function. GB2's scale parameter is
\begin{equation}
\label{betaGB2}
\beta=(\frac{\kappa_\alpha}{\kappa_2})^{2 / \alpha}
\end{equation}
and its shape parameters are $\alpha$,
\begin{equation}
\label{pGB2}
p=\frac{1}{\alpha}(-1+\alpha +\frac{2 \gamma \theta}{\kappa_\alpha^2})
\end{equation}
and
\begin{equation}
\label{qGB2}
q=\frac{1}{\alpha}(1+\frac{2 \gamma}{\kappa_2^2})
\end{equation}
The steady-state distribution of (\ref{GB2SDE}) is Generalized Inverse Gamma (GIGa) for $\kappa_{\alpha}=0$ \cite{ma2013distribution,ma2014model} and Generalized Gamma (GGa) for $\kappa_2=0$. 

For $\alpha=1$ we return to the mean-reverting -- multiplicative-Heston \cite{dashti2018combined} -- model
\begin{equation}
\mathrm{d}v = -\gamma(v - \theta )\mathrm{d}t + \sqrt{\kappa_M^2 v^2 + \kappa_H^2 v}\mathrm{d}W_t^{(2)}
\label{BPSDE}
\end{equation}
Its steady-state distribution is Beta Prime (BP)
\begin{equation}
BP(v; p,q,\beta)=\frac{ (1+({\frac{v}{\beta}}))^{-p-q}(\frac{v}{\beta})^{-1+p}}{\beta B(p,q)}
\label{BPxx}
\end{equation}
with the scale parameter
\begin{equation}
\label{betaBP}
\beta=(\frac{\kappa_H}{\kappa_M})^{2}
\end{equation}
and shape parameters,
\begin{equation}
\label{pBP}
p=\frac{2 \gamma \theta}{\kappa_H^2}
\end{equation}
and
\begin{equation}
\label{qBP}
q=1+\frac{2 \gamma}{\kappa_M^2}
\end{equation}
It is required that $p>1$, since PDF must be zero at $v=0$. (This condition also assures that the distribution has a bell shape.) We also require that $q>2$, that is $\frac{2 \gamma}{\kappa_M^2}>1$ which assures that variance exists. For multiplicative model, $\kappa_H = 0$, the steady-state distribution of (\ref{BPSDE}) is Inverse Gamma (IGa) and for Heston model (Cox-Ingersoll-Ross model of volatility), $\kappa_M = 0$, it is Gamma (Ga) \cite{praetz1972distribution,nelson1990arch, heston1993closed, dragulescu2002probability,liu2017distributions}. 

In this Section we will consider "reduced" covariance $cov[v_tv_{t+\tau}]/<{v_t}>^2$, that is $cov[v_tv_{t+\tau}]/\theta^2$ (compare with $corr[v_tv_{t+\tau}]$ (\ref{corrvt})). The reason is that we want to use the market data to determine model parameters. In what follows, the discussion will be limited to the mean-reverting models. Using (\ref{meanVarCor}) and {\ref{varvt}}, we find for the multiplicative-Heston model
\begin{equation}
\frac{cov[v_t v_{t+\tau}]}{<{v_t}>^2}=\frac{\kappa_M^2\theta^2+\kappa_H^2\theta}{2\gamma-\kappa_M^2} exp(- \gamma \tau)
\label{covreducedHM}
\end{equation}
The result for multiplicative and Heston models can be recovered by setting $\kappa_H = 0$ and $\kappa_M = 0$ respectively:
\begin{equation}
\frac{cov[v_t v_{t+\tau}]_M}{<{v_t}>^2}=\frac{\kappa_M^2\theta^2}{2\gamma-\kappa_M^2} exp(- \gamma \tau)
\label{covreducedM}
\end{equation}
\begin{equation}
\frac{cov[v_t v_{t+\tau}]_H}{<{v_t}>^2}=\frac{\kappa_H^2\theta}{2\gamma} exp(- \gamma \tau)
\label{covreducedH}
\end{equation}

To find leverage, we use (\ref{LeverageSimple}). For multiplicative-Heston model we find 
\begin{equation}
\begin{split}
& \mathscr{L}_{MH}(\tau) = \frac{\kappa _M \left(\frac{\kappa _H^2}{\kappa _M^2}\right){}^{3/2} B\left(\frac{2 \gamma  \theta }{\kappa _H^2}+1,\frac{2 \gamma }{\kappa _M^2}-\frac{1}{2}\right)}{\theta ^2 B\left(\frac{2 \gamma  \theta }{\kappa _H^2},\frac{2 \gamma }{\kappa _M^2}+1\right)}\end{split} e^{-\gamma\tau}
\label{LMH}
\end{equation}
The result for multiplicative and Heston models can be recovered by setting $\kappa_H = 0$ and $\kappa_M = 0$ respectively or by calculating directly with (\ref{LeverageSimple}) (for Heston model, see also \cite{perello2003random}). We find
\begin{equation}
\mathscr{L}_{M}(\tau)= \frac{\rho\kappa_{M} (\frac{2 \gamma}{\kappa_{M}^2})^{\frac{1}{2}} \Gamma(\frac{2 \gamma}{\kappa_{M}^2}-\frac{1}{2})}{ \theta^{\frac{1}{2}} \Gamma(\frac{2 \gamma}{\kappa_{M}^2})} e^{-\gamma\tau}
\label{LM}
\end{equation}
where $\Gamma$ is the gamma function, and 
\begin{equation}
\mathscr{L}_{H}(\tau)=\frac{\rho \kappa_H}{\theta}e^{-\gamma\tau}
\label{LH}
\end{equation}
for multiplicative and Heston model respectively. 
\subsection{Numerical Fitting \label{Numerical}}
\begin{figure}[!htbp]
\centering
\begin{tabular}{cc}
\includegraphics[width = 0.49 \textwidth]{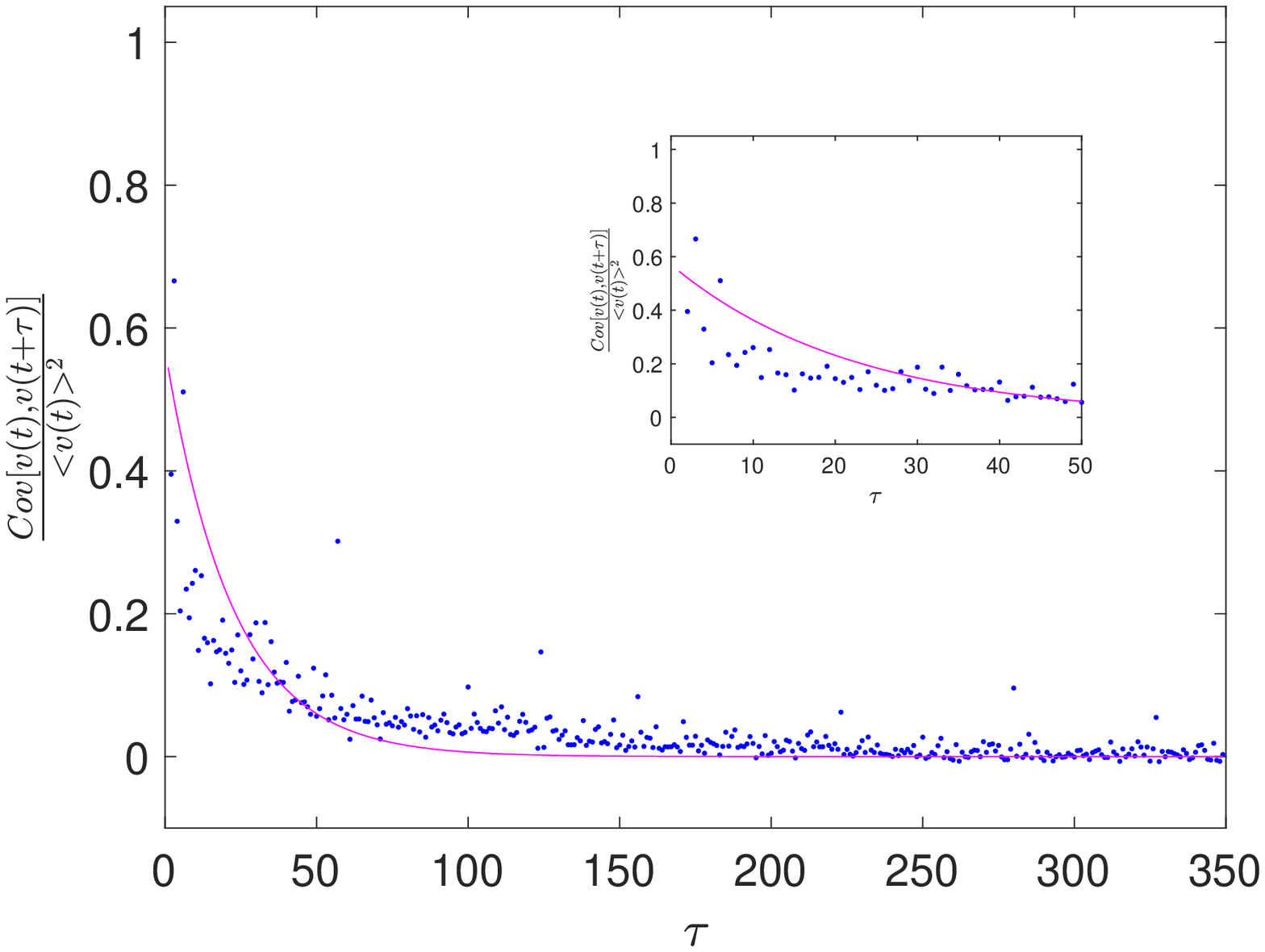}
\includegraphics[width = 0.49 \textwidth]{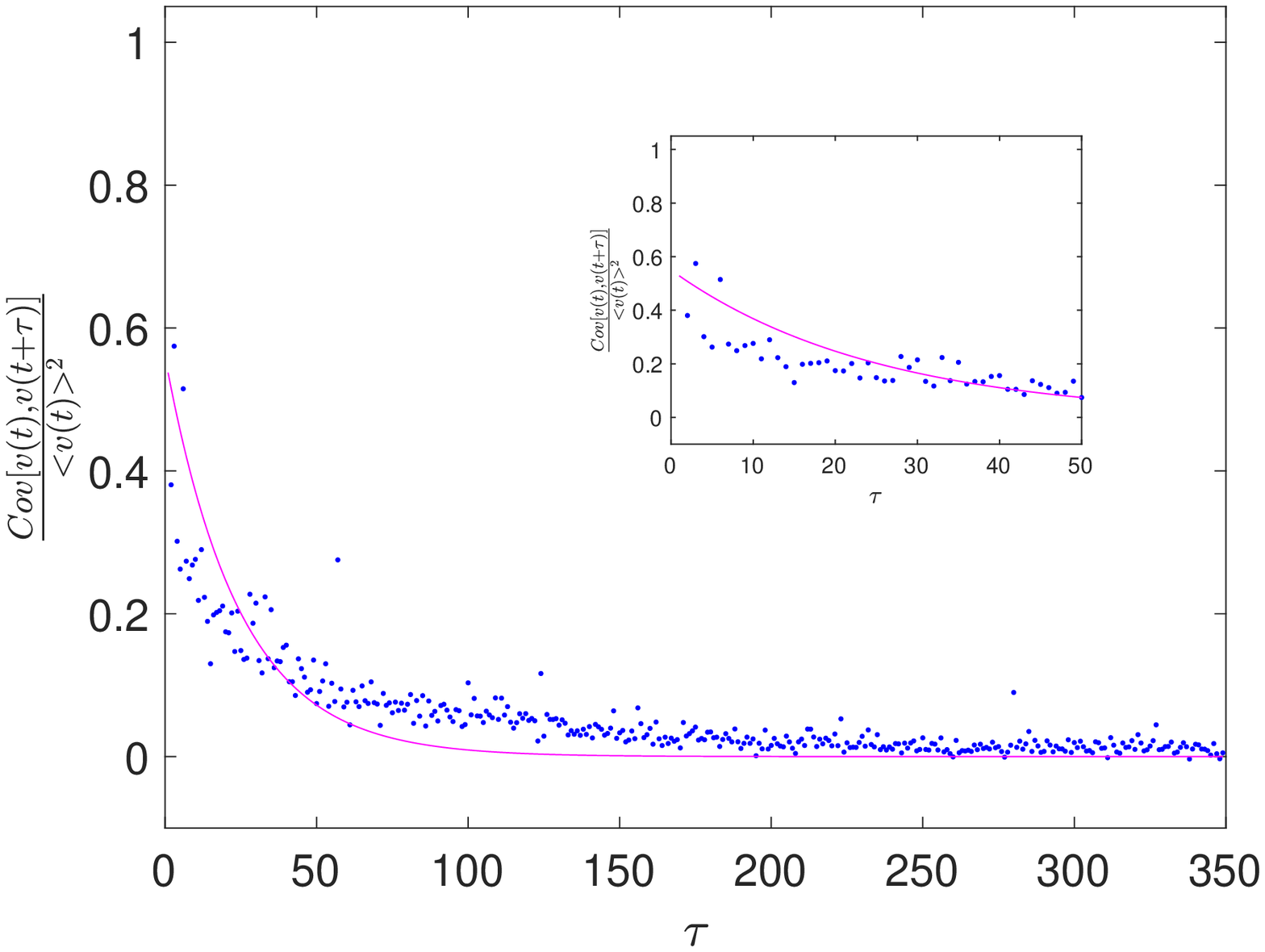} \\
\includegraphics[width = 0.49 \textwidth]{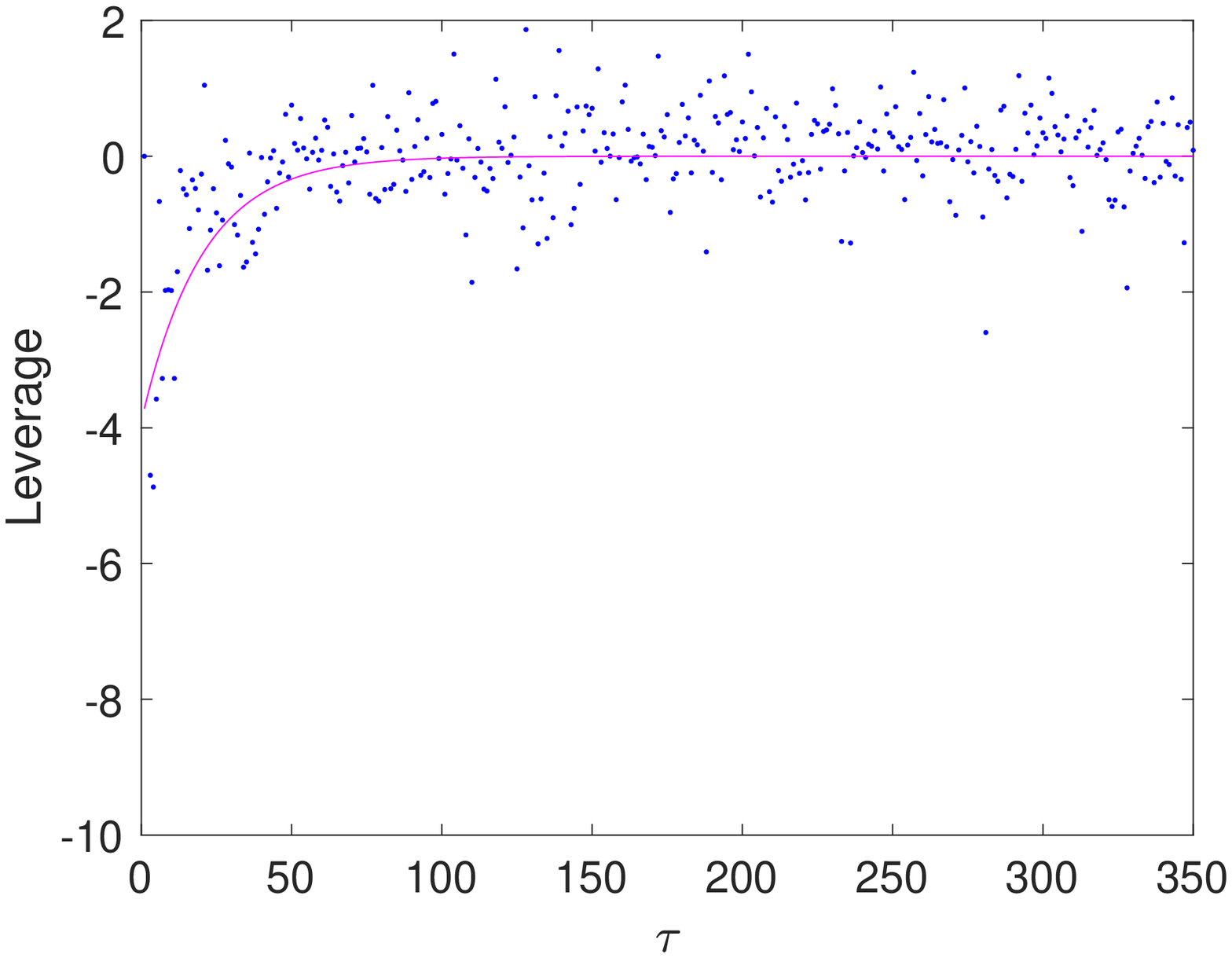}
\includegraphics[width = 0.49 \textwidth]{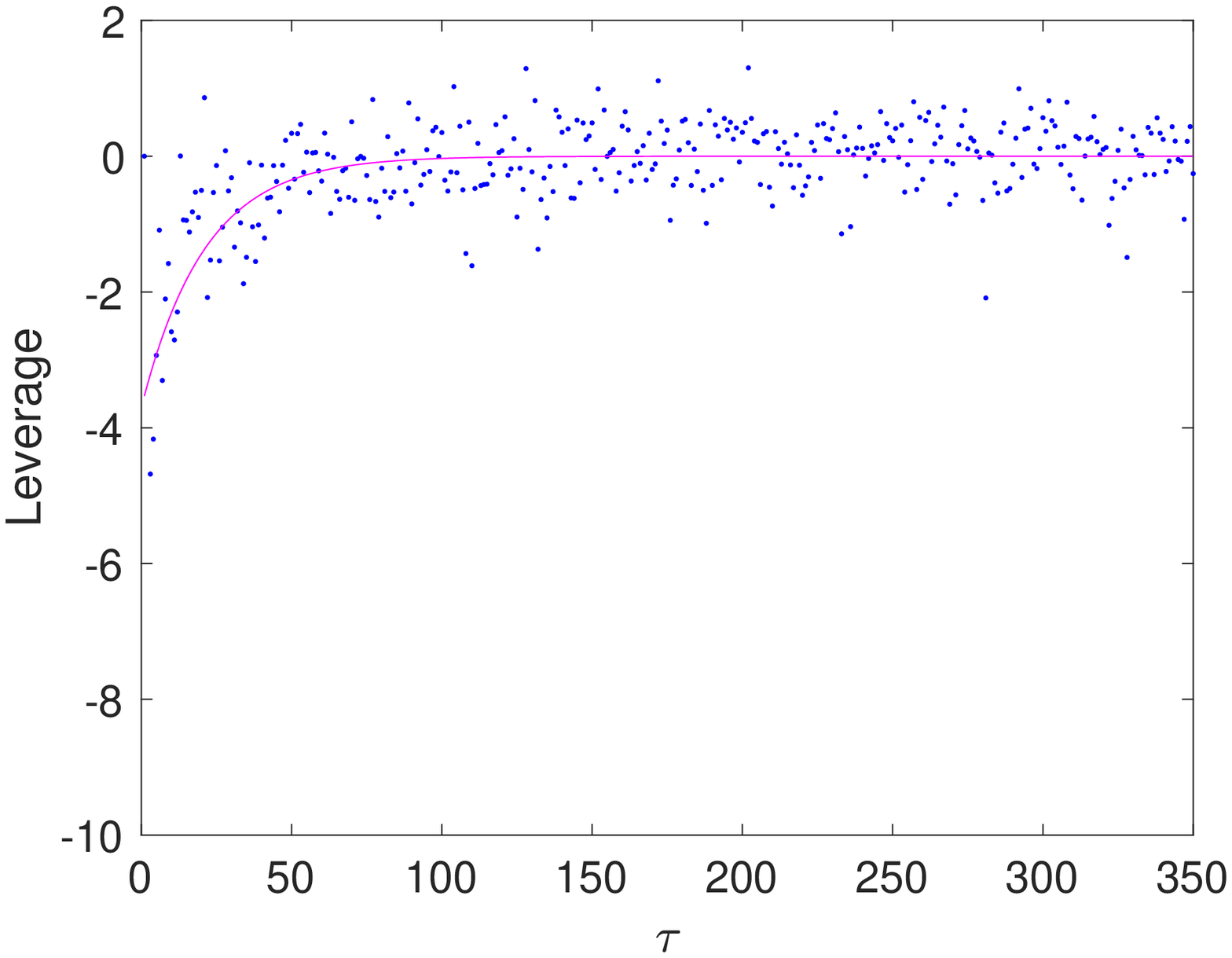}
\end{tabular}
\caption{Top row: $\frac{cov[v_tv_{t+\tau}]}{<{v_t}>^2}$. Bottom row: Leverage (\ref{LeverageDef}). Left column: DJIA. Right column: S\&P500.}
\label{fits}
\end{figure}

We use market data for daily returns. For our numerical fitting we adopt the following procedure:
\begin{enumerate}
\item We use $\gamma$ obtained in Sec. \ref{Variance};
\item We use (\ref{expxt2}) to obtain $<v_t>$ ( that is, $\theta$);
\item We use the second of (\ref{dxt2corrmulti}) to obtain $cov[v_t v_{t+\theta}]= <v_t v_{t+\theta}>-<v_t>^2$;
\item We fit $cov[v_t v_{t+\theta}]/<v_t>^2$ with $A exp(-\gamma \tau)$ to determine $A$;
\item We use (\ref{covreducedM}) and (\ref{covreducedH}) to determine $\kappa_M$ and $\kappa_H$ respectively;
\item We use $\kappa_M$ and $\kappa_H$ obtained in previous step and (\ref{LM}) and (\ref{LH}) to find $\rho$ and $\gamma_L$, relaxation parameter found from leverage.
\end{enumerate}
Fig. \ref{fits} shows plots of $cov[v_t v_{t+\theta}]/<v_t>^2$ and leverage and their fits and the results of the above fitting procedure are summarized in Table \ref{params}. Notice that we use only multiplicative and Heston models since for the combined multiplicative-Heston model we can not independently find $\kappa_M$ and $\kappa_H$ using this procedure. However, we can determine those for the combined multiplicative-Heston model as a function of the number of days of returns, beginning with daily returns, using the stocks returns distribution function associated with this model and its BP steady-state distribution \cite{dashti2018combined}.

\begin{table}[!htb]
\label{params}
  \begin{minipage}{.5\textwidth}
\begin{center}
\begin{tabular}{ c c } 
\multicolumn{2}{c}{DJIA Parameters} \\
 \hline
   Parameters &  \\ 
 \hline
 $\gamma$ & $0.045$\\ 
 \hline
 $\theta$ & $9.52 \times 10^{-5}$ \\ 
 \hline
  $\kappa_{M}$ & $0.24$ \\ 
 \hline
 $\kappa_{H}$ & $2.18 \times 10^{-3}$\\ 
 \hline
  $\rho_{M}$ & $-0.114$\\
 \hline
 $\rho_{H}$ & $-0.165$\\
 \hline
   $\gamma_{L}$ & $0.049 $\\
 \hline
\end{tabular}
\end{center}
 \end{minipage}
   \begin{minipage}{.5\textwidth}
\begin{center}
\begin{tabular}{ c c } 
\multicolumn{2}{c}{S\&P Parameters} \\
 \hline
   Parameters &  \\ 
 \hline
  $\gamma$ & $0.041$ \\ 
 \hline
 $\theta$ & $9.81 \times 10^{-5}$ \\ 
 \hline
 $\kappa_{M}$ & $0.22$ \\ 
 \hline
 $\kappa_{H}$ & $2.17\times 10^{-3}$ \\ 
 \hline
  $\rho_{M}$ & $-0.123$\\
 \hline
 $\rho_{H}$ & $-0.162$\\
 \hline
  $\gamma_{L}$ & $0.047 $\\
 \hline
\end{tabular}
\end{center}
  \end{minipage}
\end{table}

\begin{figure}[!htbp]
\centering
\begin{tabular}{cc}
\includegraphics[width = 0.4 \textwidth]{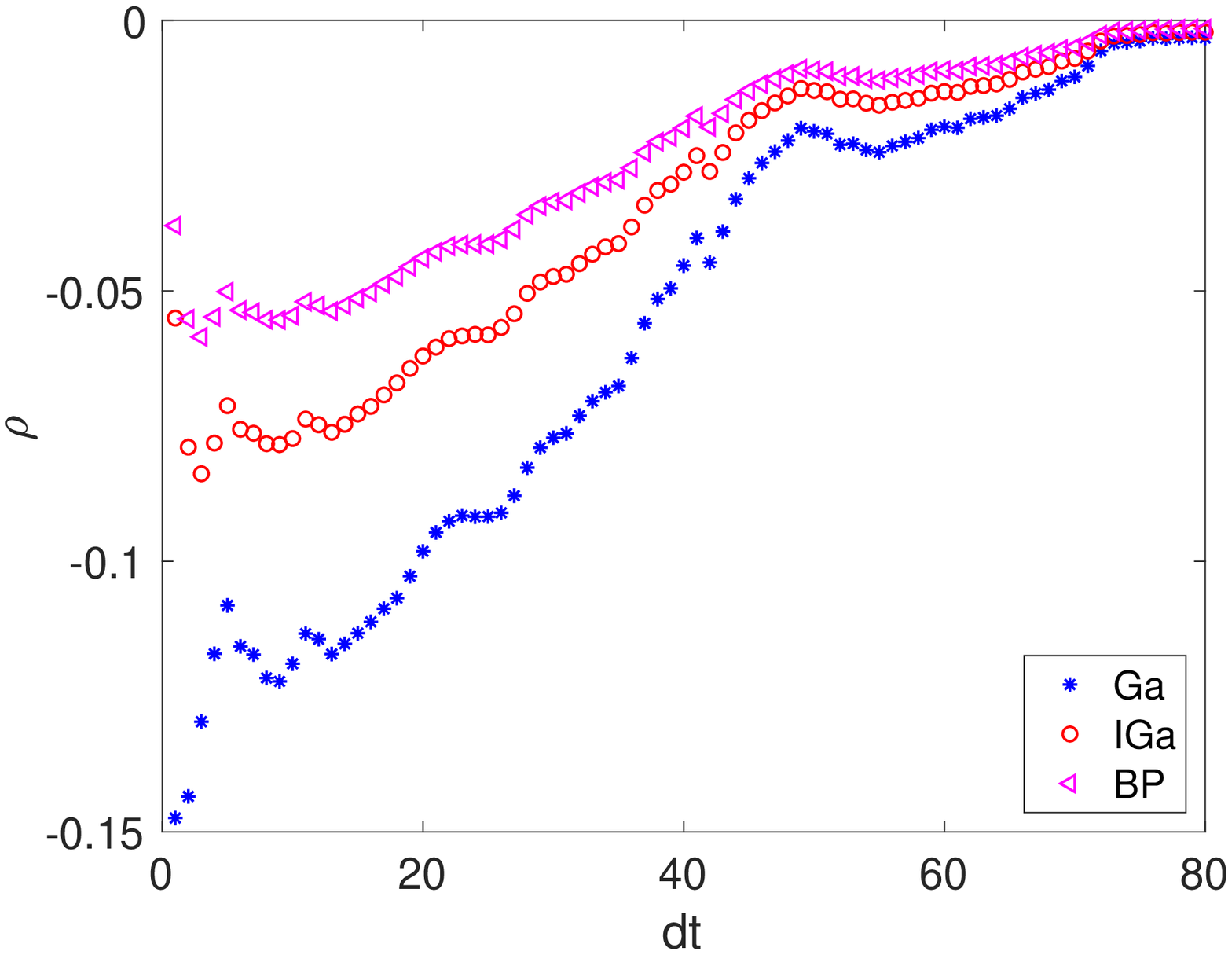}
\includegraphics[width = 0.4 \textwidth]{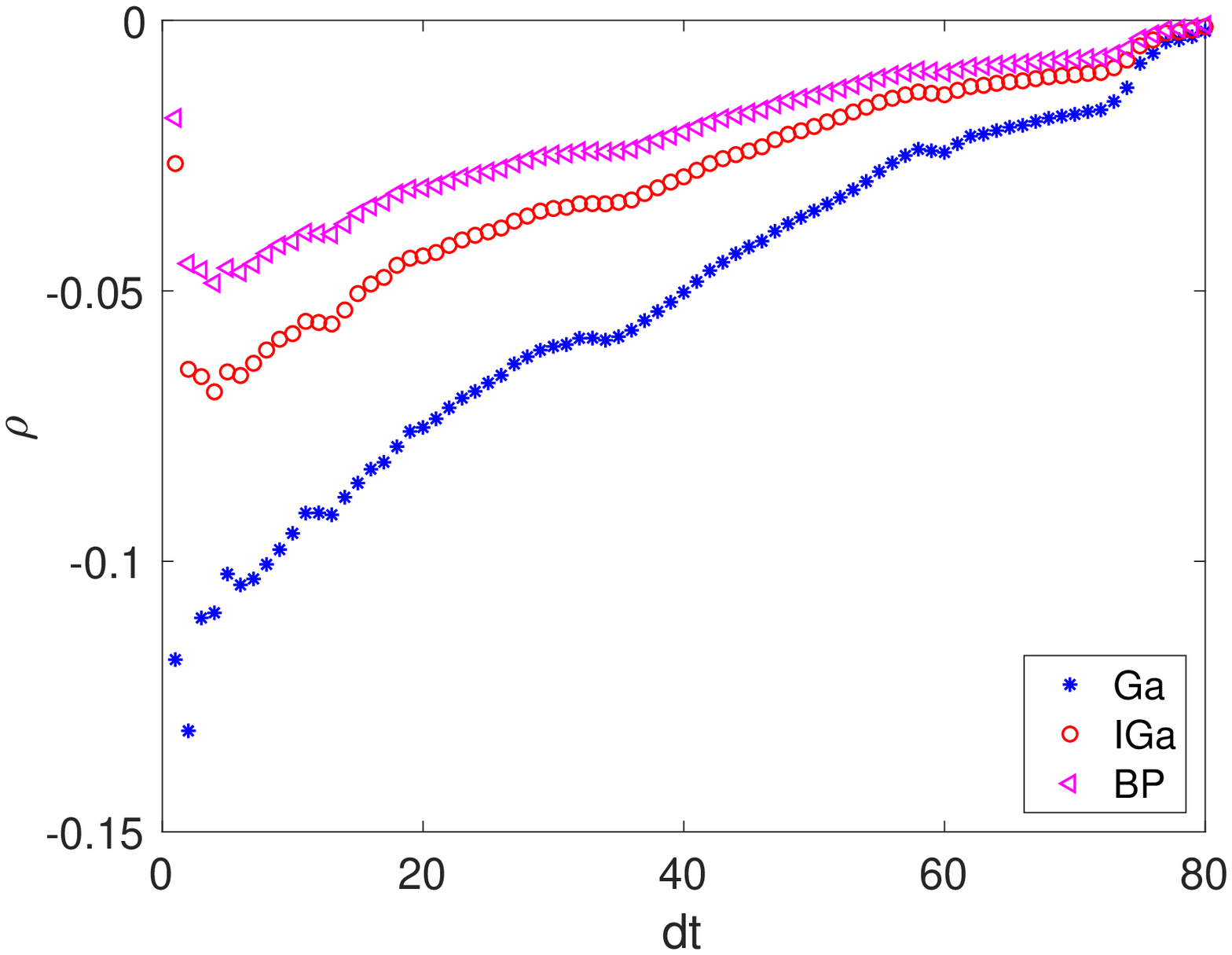}
\end{tabular}
\caption{Cross-correlation $\rho$ for multi-day returns as a function of days of accumulation. Left: DJIA. Right: S\&P500.} 
\label{rhomulti}
\end{figure}

We also conducted a study of the leverage for multi-day returns. Towards this end we used (\ref{LMH})-(\ref{LH}) with the values of $\gamma$, $\theta$,  $\kappa_M$ and $\kappa_H$ obtained in \cite{dashti2018combined}. The results for cross-correlation $\rho$ are shown in Fig. \ref{rhomulti}. Obviously, $\rho$ decays rapidly with the number of days of accumulation of returns.

\section{Conclusions\label{Conclusions}}

We found that the correlation function (Pearson correlation coefficient) of stochastic variance (\ref{corrvt}) in mean-reverting models depends only on one - relaxation -- parameter and that the variance of the variance can be found from a general, model independent formula (\ref{varvt}). We also argued that leverage can be  found from a general, model independent formula (\ref{LeverageSimple}). 

We investigated the relationship between the correlation functions of stochastic variance and of realized variance for multi-day returns, (\ref{dxt2corrmulti}), and investigated the latter as a function of the number of days of accumulation of returns. We also empirically investigated the correlation function of realized variance and showed that it can be described by a single-parameter exponent, with inverse time parameter itself exponentially decreasing to its finite asymptotic values with the increase of the number of days of returns accumulation. We were unable to propose an explanation of this behavior. 

For two specific volatility models -- multiplicative and Heston -- we used the correlation function and leverage to determine model parameters and cross-correlation between stochastic volatility and stock returns. We also showed that cross-correlation decays rapidly for multi-day returns. 

We examined correlations and relaxation specifically for Heston model and showed that it displays a progression of relaxation times that are reflected in cumulants' relaxation. Finally, we proposed that the distribution of relaxation times is best described by an Inverse Gaussian.

\appendix
\section{Correlation Function of Multi-day Realized Variance \label{corrdxt2}}
In Section \ref{CorrLev} we argued that for the mean-reverting models, (\ref{general}), the correlation function of stochastic variance is given, per (\ref{corrvt}), by $corr[v_t v_{t+\tau}] =e^{-\gamma \tau}$. The problem with this prediction is relating $corr[v_t v_{t+\tau}]$ to actual market quantities. For daily returns it is given by the l.h.s. of (\ref{daily}), however its fit with the $e^{-\gamma \tau}$ in Fig. \ref{Correlation} is rather poor. The latter may be attributed to that continuous, mean-reverting models of stochastic variance are not a good match for daily returns. On the other hand, they may be more appropriate for multi-day returns \cite{dashti2018combined,liu2017distributions}. Unfortunately, the relationship between multi-day realized \footnote{Notice that multi-day realized variance, that is variance calculated for multi-day accumulation of stock returns, is different from realized variance related to realized volatility, which is calculated by addition of daily realized variances \cite{dashti2018realized}.} and stochastic variances is complicated, per (\ref{dxt2corrmulti}), and so is determining dependence on $\tau$ in presence of two time scales, $t$ and $\gamma$. 

\begin{figure}[!htbp]
\centering
\begin{tabular}{cc}
\includegraphics[width = 0.49 \textwidth]{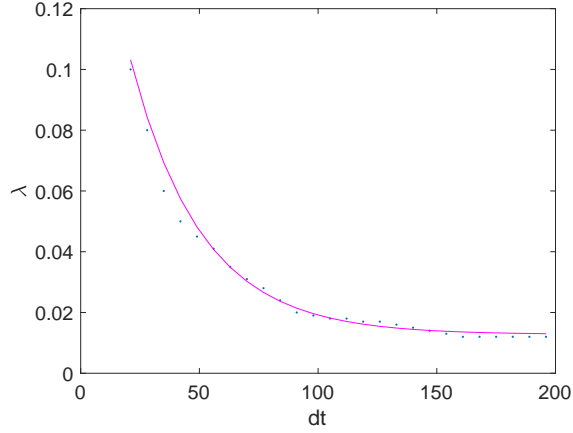} 
\end{tabular}
\caption{$a + (b-a) exp^{- \lambda t}$ best fit of $\gamma_1$ as function days of accumulation, beginning with monthly returns (21 days), with $a = 0.0128, b = 0.182, \lambda = 0.0335$ and $r^2 = 0.9926$.} 
\label{expfit}
\end{figure}

\begin{figure}[!htbp]
\centering
\begin{tabular}{cc}
\includegraphics[width = 0.49 \textwidth]{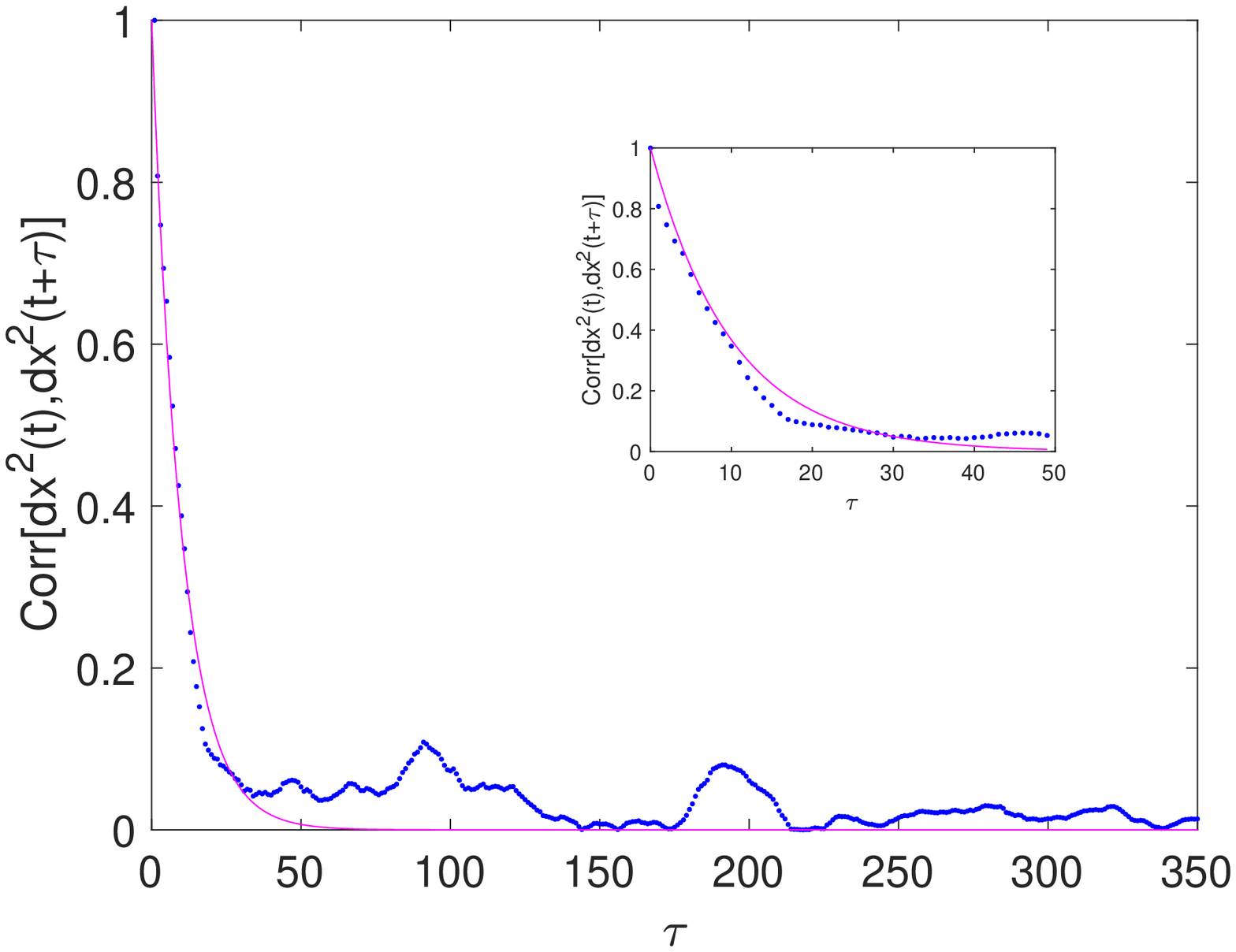}
\includegraphics[width = 0.49 \textwidth]{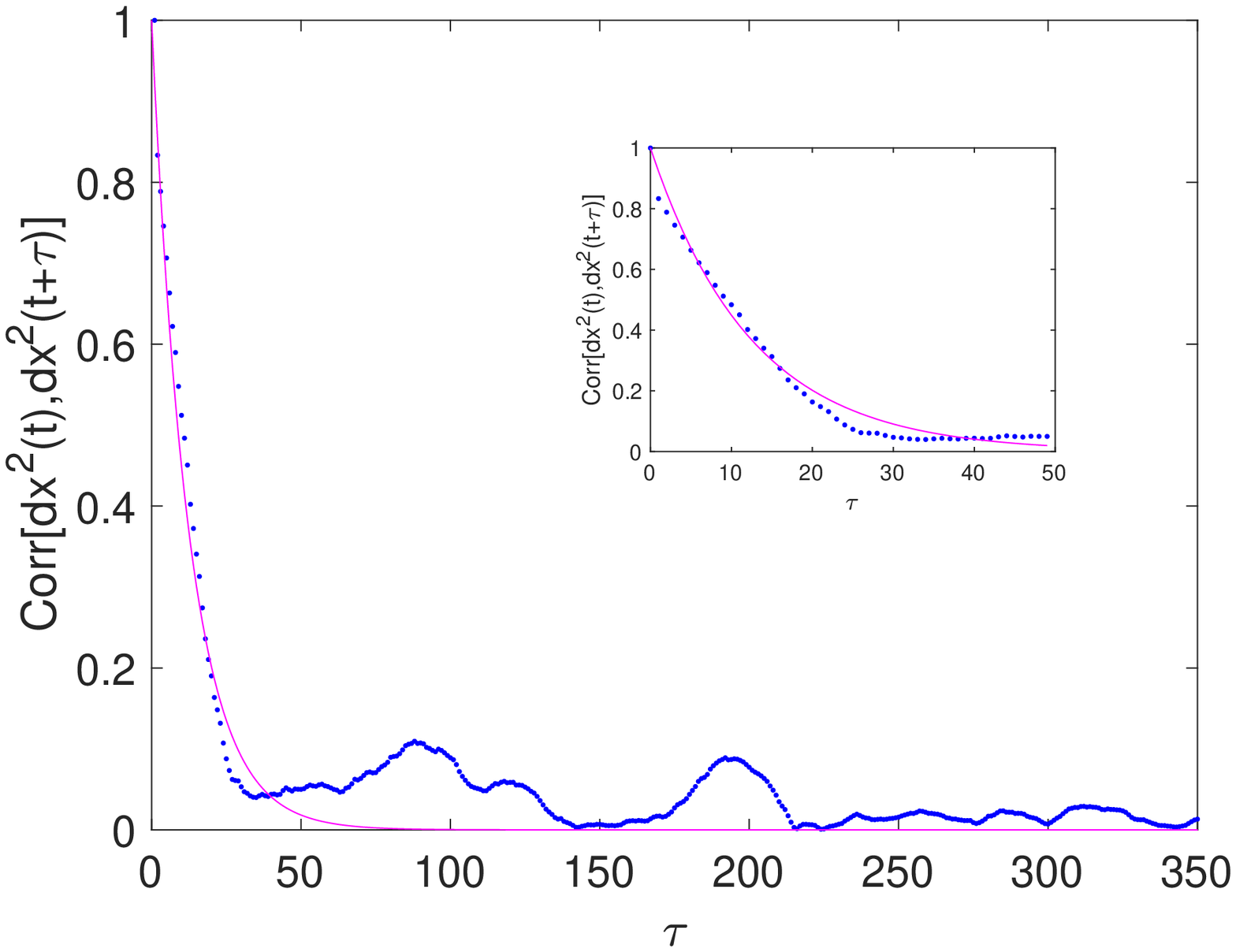} \\
\includegraphics[width = 0.49 \textwidth]{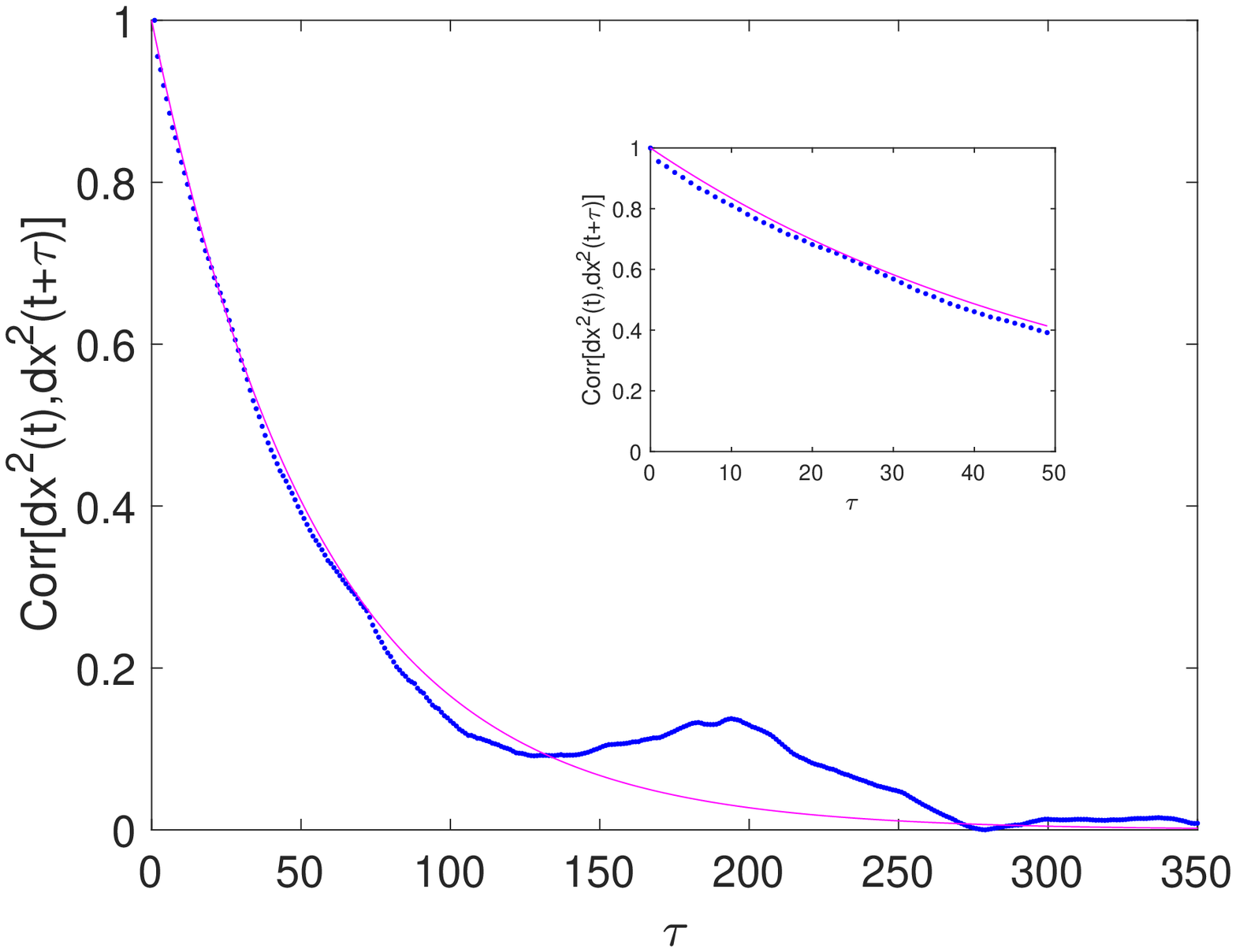}
\includegraphics[width = 0.49 \textwidth]{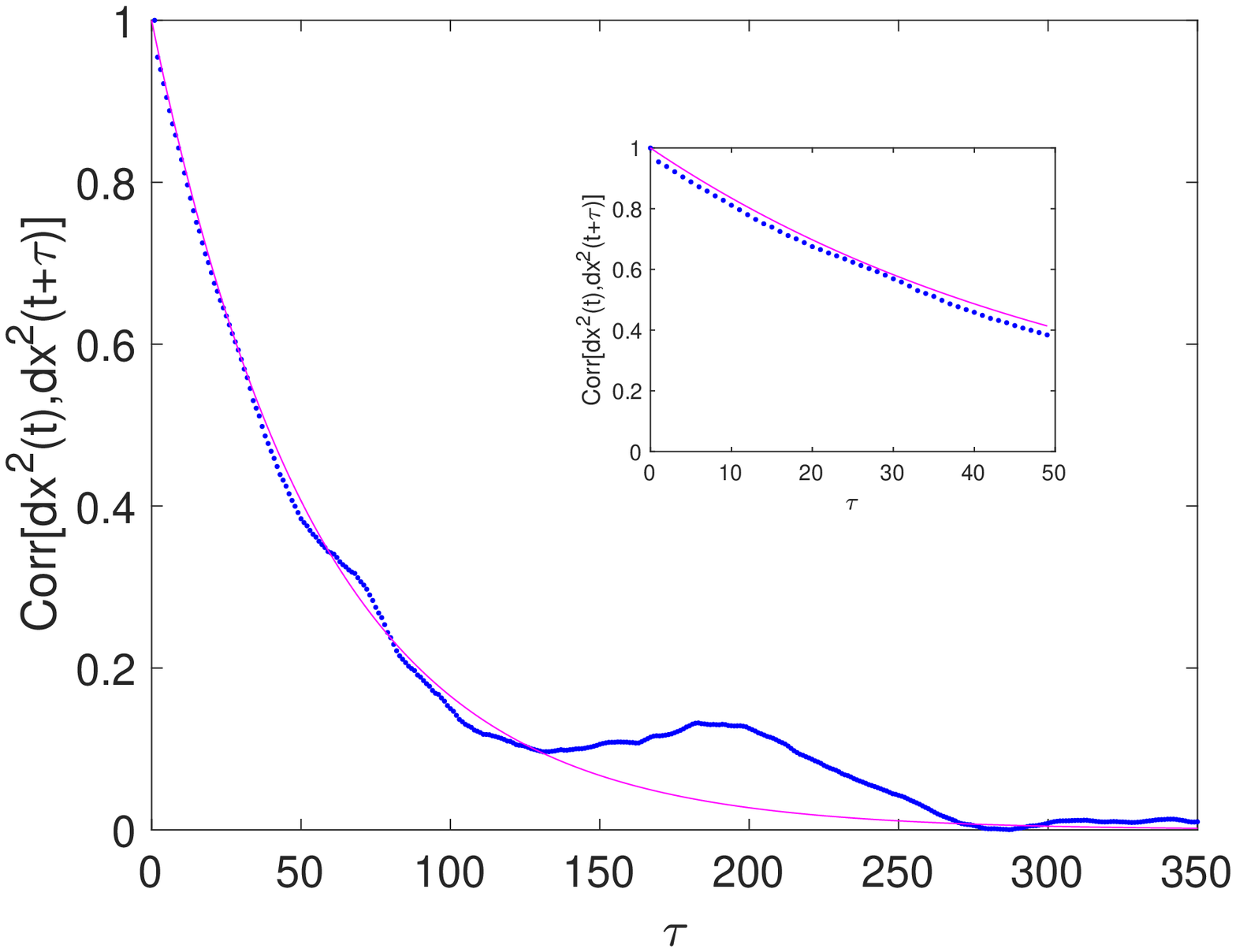} \\
\includegraphics[width = 0.49 \textwidth]{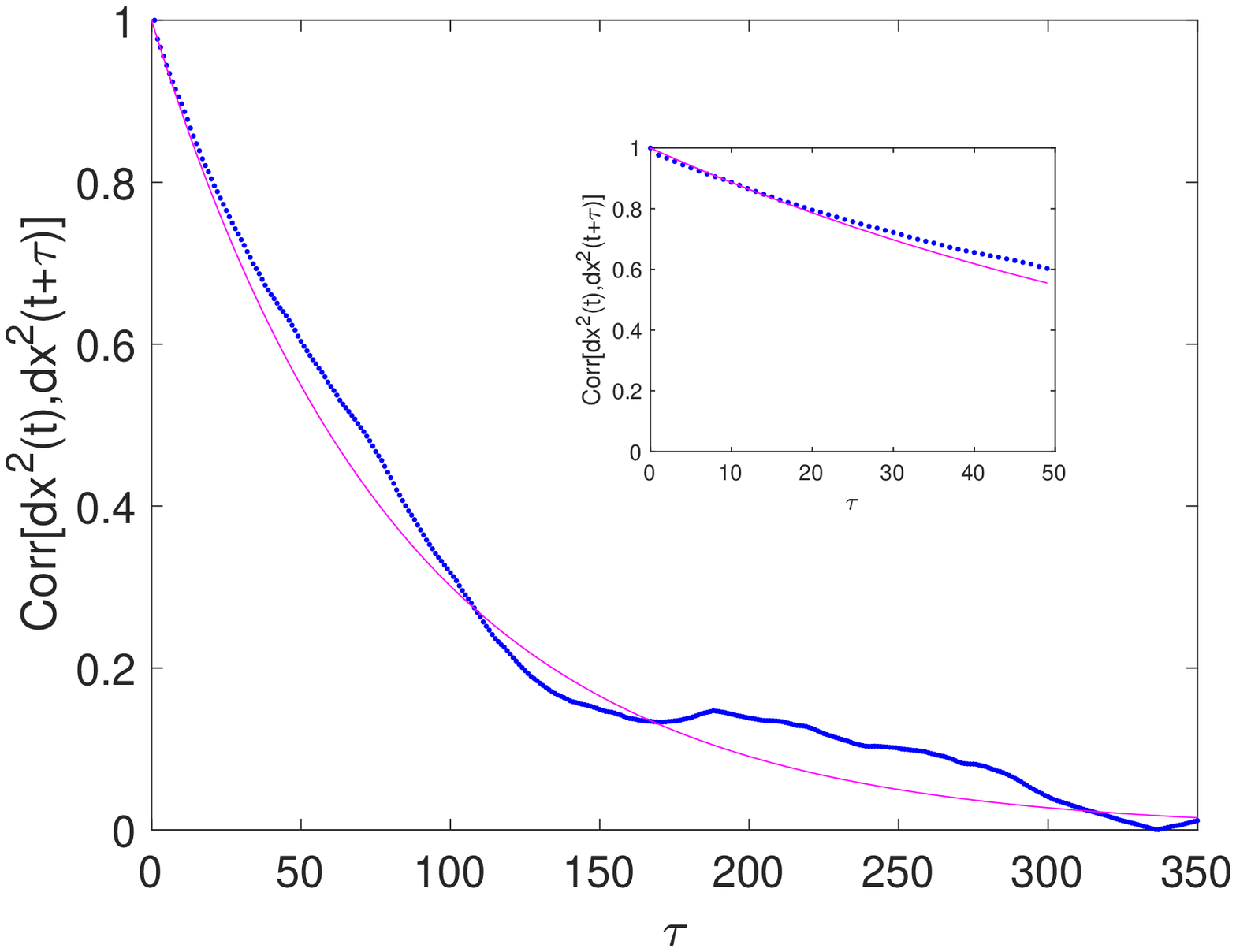}
\includegraphics[width = 0.49 \textwidth]{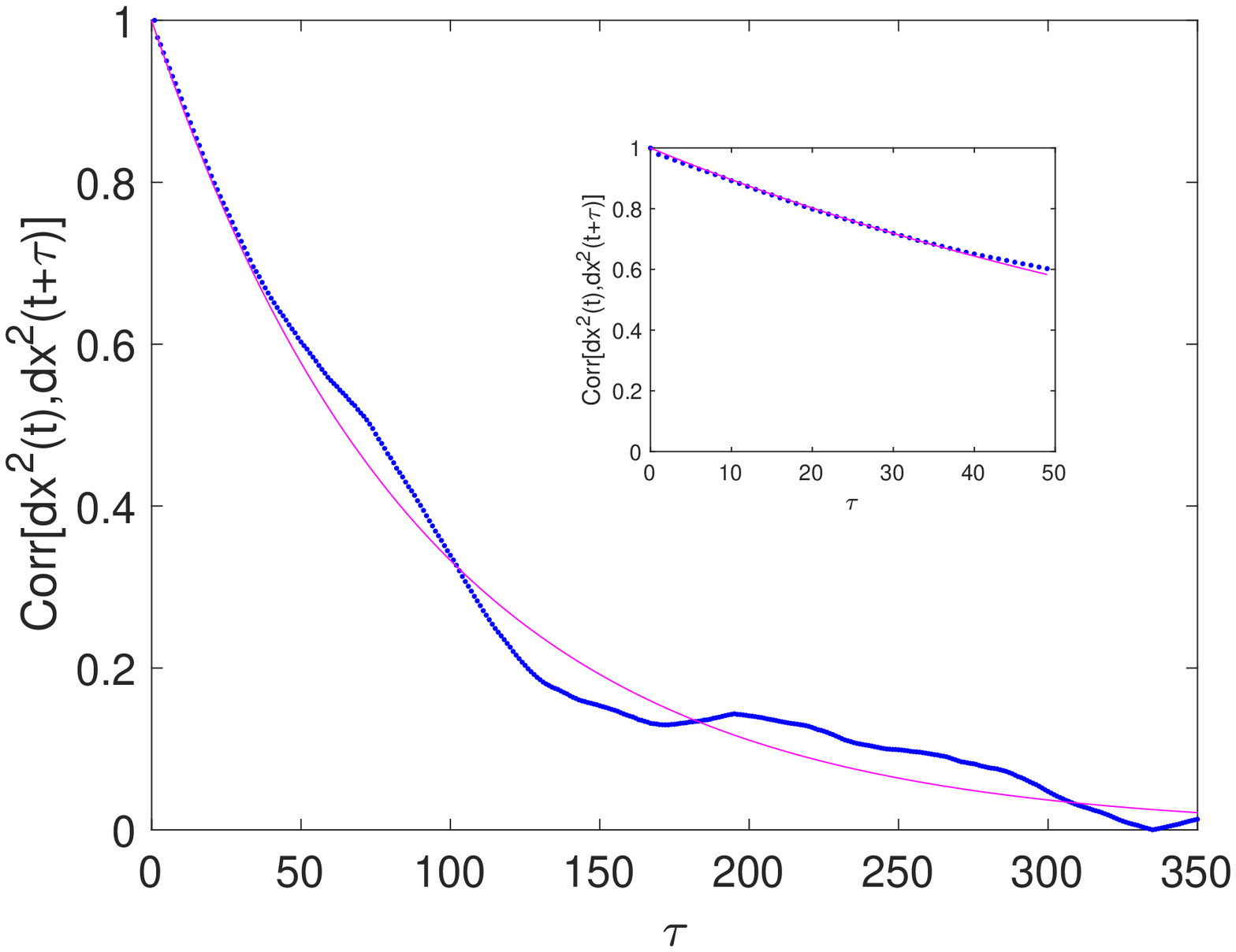} 
\end{tabular}
\caption{sample $a e^{-\gamma_1 \tau}$ fits for 21, 28 (top row), 105, 112 (middle row), 189 and 196 (bottom row) days of accumulation with parameter values from Table \ref{gamma1}.} 
\label{corrdxt2fits}
\end{figure}

Consequently, we surmised, per(\ref{multiday}), that the correlation function of multi-day realized variance is expressed by a pure exponent, $corr[\mathrm{d}x_t^2 \mathrm{d}x_{t+\tau}^2] \approx e^{-\gamma_1 \tau}$. We fitted $corr[\mathrm{d}x_t^2 \mathrm{d}x_{t+\tau}^2]$ with $a e^{-\gamma_1 \tau}$ for multi-day S\&P returns (DJIA is very similar) and summarized our empirical findings, including $r^2$ statistics, in Table \ref{gamma1}
\begin{table}[!htbp]
\centering
\caption{Fitting result}
\label{gamma1}
\begin{tabular}{cccc} 
\hline
 t &    a &  $\gamma_1$ & $r^2$ \\
 \hline
 7 & 0.57 & 0.09 & 0.71\\
 14 & 0.80 & 0.11 & 0.75\\
 21 & 0.95 & 0.10 & 0.78\\
 28 & 0.97 & 0.08 & 0.80\\
 35 & 0.96 & 0.06 & 0.84\\
 42 & 0.96 & 0.05 & 0.87\\
 49 & 0.98 & 0.045 & 0.89\\
 56 & 0.99 & 0.041 & 0.90\\
 63 & 0.99 & 0.035 & 0.92\\
 70 & 0.99 & 0.031 & 0.93\\
 77 &    1.00 & 0.028 & 0.93\\
 84 & 1.00 & 0.024 & 0.94\\
 91 & 1.00 & 0.020 & 0.94\\
 98 & 1.00 & 0.019 & 0.95\\
 105 & 1.00 & 0.018 & 0.95\\
 112 & 1.00 & 0.018 & 0.96\\
 119 & 1.00 & 0.017 & 0.96\\
 126 & 1.00 & 0.017 & 0.97\\
 133 & 1.00 & 0.016 & 0.97\\
 140 & 1.00 & 0.015 & 0.98\\
 147 & 1.00 & 0.014& 0.98\\
 154 & 1.00 & 0.013 & 0.98\\
 161 & 1.00 & 0.012 & 0.98\\
 168 & 1.00 & 0.012& 0.98\\
 175 & 1.00 & 0.012 & 0.97\\
 182 & 1.00 & 0.012 & 0.97\\
 189 & 1.00 & 0.012 & 0.98\\
 196 & 1.00 & 0.012 & 0.98\\
 \hline
 \hline
\end{tabular}
\end{table}
In Fig. \ref{expfit} we show exponential fit of $\gamma_1$ itself as a function of days of accumulation. In Fig. \ref{corrdxt2fits} we show sample $a e^{-\gamma_1 \tau}$ fits for 21, 28, 105, 112, 189 and 196 days of accumulation with parameter values from Table \ref{gamma1}. At present, we do not have plausible interpretation of our empirical findings.

\section{Correlations and Relaxation in Heston (Cox-Ingersoll-Ross) Model \label{CIRmodel}}
\subsection{Eigenvalue Solution of Fokker-Planck Equation \label{FPapproach}}
We have previously investigated correlations and relaxation in multiplicative model \cite{liu2017correlation}. Here we will apply the same approach to Heston (Cox-Ingersoll-Ross) model. To remain consistent with notations of \cite{liu2017correlation}, we replace $v_t$ with $x$ (not to confuse with stock returns), drop superfluous indices and write the model as 
\begin{equation}
\mathrm{d}x = -\gamma(x - \theta)\mathrm{d}t + \kappa \sqrt{x}\mathrm{d}W
\label{CIR}
\end{equation}
Obviously, via rescaling $x/\theta \rightarrow x$ and $\kappa \sqrt{\theta} \rightarrow  \kappa$, this equation can be reduced to that with the unity mean
\begin{equation}
\mathrm{d}x = -\gamma(x -1)\mathrm{d}t + \kappa \sqrt{x}\mathrm{d}W
\label{CIR1}
\end{equation}
For now, however, we will proceed with (\ref{CIR}). 

The Fokker-Planck equation for this process is given by 
 \begin{equation}
\frac{\partial P(x,t)}{\partial t}=\gamma \frac{\partial (x-\theta) P(x,t)}{\partial x}+\frac{\kappa^2}{2}\frac{\partial^2 x P(x,t)}{\partial x^2}
\label{FPeq}
\end{equation}
To find correlations and relaxation, we use an eigenvalue approach \cite{schenzle1979multiplicative} to solving it. Namely, we seek the solution in the following form:
\begin{equation}
P(x,t)=P_0 (x)+P(\lambda ; x)e^{-\lambda t}
\label{Problemsoleigen}
\end{equation}
where $\lambda>0$ and $P_0(x)$ is a Ga steady-state distribution of (\ref{CIR})
\begin{equation}
P_0(x)=\frac{e^{-\frac{2\gamma x}{\kappa^2} }{(\frac{2\gamma x}{\kappa^2})}^{\frac{2\gamma\theta}{\kappa^2}-1}}{\frac{\kappa^2}{2\gamma}\Gamma(\frac{2\gamma\theta}{\kappa^2})}, \hspace{.21cm} \frac{2\gamma\theta}{\kappa^2} > 1
\label{SSrelax}
\end{equation}
where the latter assures that $P_0(0)=0$. $P(\lambda ; x)e^{-\lambda t}$ describe relaxation to the steady state and we should also have $P(\lambda ; 0)=0$

Substitution of (\ref{Problemsoleigen}) into (\ref{FPeq}) yields
\begin{equation}
\frac{\kappa^2}{2}(x P(\lambda ; x))^{\prime\prime}+\gamma((x-\theta)P(\lambda ; x))^{\prime}+\lambda P(\lambda ; x)=0
\label{solveeigen}
\end{equation}
which has two solutions
 \begin{equation}
P_1(\lambda ; x) \propto e^{-\frac{2x\gamma}{\kappa^2}}U(1-\frac{2\gamma \theta}{\kappa^2}-\frac{\lambda}{\gamma},2-\frac{2\gamma\theta}{\kappa^2},\frac{2x\gamma}{\kappa^2})
\label{solutioneigen1}
\end{equation}
and
 \begin{equation}
P_2(\lambda ; x) \propto e^{-\frac{2x\gamma}{\kappa^2}}    {\mathbf{L}}^{(1-\frac{2\gamma\theta}{\kappa^2})}_{(\frac{2\gamma \theta}{\kappa^2}+\frac{\lambda}{\gamma}-1)}(\frac{2x\gamma}{\kappa^2})
\label{solutioneigen2}
\end{equation}
where $U$ is Tricomi's confluent hypergeometric function and L is Laguerre polynomial function. Condition $P(\lambda ; 0)=0$ cannot be satisfied by $P_2(\lambda ; x)$ and for $P_1(\lambda ; x)$ it leads to quantization of $\lambda$, $\lambda_n=n\gamma$, where $n>0$ is an integer. Consequently, the eigenfunctions of (\ref{solveeigen}) are given by 
\begin{equation}
P_1(\lambda_n ; x)\equiv P_n(x)\propto e^{-\frac{2x\gamma}{\kappa^2}}U(1-\frac{2\gamma \theta}{\kappa^2}-n,2-\frac{2\gamma\theta}{\kappa^2},\frac{2x\gamma}{\kappa^2}), \hspace{.21cm} \lambda_n=n\gamma
\label{solutioneigenwithm}
\end{equation}

The correlation function can be found as  \cite{schenzle1979multiplicative}
\begin{equation}
<\delta x(t+\tau)\delta x(t)>=\sum_{n} g_n^2 e^{-\lambda_n \tau}
\label{generalcorfun}
\end{equation}
where
 \begin{equation}
g_n \propto \int xP(\lambda_n;x)\mathrm{d}x=\int \delta xP(\lambda_n;x)\mathrm{d}x
\label{gfunc}
\end{equation}
Using (\ref{solutioneigenwithm}), we find 
\begin{equation}
g_n \propto \frac{\kappa^4\Gamma(1+\frac{2\gamma\theta}{\kappa^2})}{4\gamma^2\Gamma(2-n)}
\label{gfunc2}
\end{equation}
Clearly, the only non-zero $g_n$ is $g_1$. Using normalization condition \cite{schenzle1979multiplicative}
\begin{equation}
\int_{0}^{\infty} \frac{P_1^2(\lambda_1;x)}{P_0(x)}\mathrm{d}x=1
\label{normalizeProb}
\end{equation}
We find
 \begin{equation}
P_1(\lambda_1;x)=\frac{e^{-\frac{2x\gamma}{\kappa^2}}(x-\theta)(\frac{2x\gamma}{\kappa^2})^{1+\frac{2 \theta \gamma}{\kappa^2}}}{x^2 \sqrt{\Gamma(1+\frac{2\gamma\theta}{\kappa^2})}\sqrt{\Gamma(\frac{2\gamma\theta}{\kappa^2})}}
\label{normalizedeigenfun}
\end{equation}
so that
\begin{equation}
g_1= \int_{0}^{\infty} P_1(\lambda_1;x) x  \mathrm{d}x=\sqrt{\frac{\theta\kappa^2}{2\gamma}}
\label{normalizeProb2}
\end{equation}
and
\begin{equation}
<\delta x(t+\tau)\delta x(t)>=\theta^2+ \frac{\theta\kappa^2}{2\gamma}e^{-\gamma\tau}
\label{corfunstart}
\end{equation}
that is \begin{equation}
\frac{<\delta x(t+\tau)\delta x(t)>-<\delta x(t)>^2}{<\delta x(t)>^2}= \frac{\kappa^2}{2\gamma\theta}e^{-\gamma\tau}
\label{corfun}
\end{equation}
which is the same result as we already found in (\ref{covreducedH}). The value of eigenvalue approach, however, is to establish multiple relaxation (time) scales, which we address next.

\subsection{Cumulant Relaxation\label{CumulRelax}}
\begin{figure}[!htbp]
\centering
\begin{tabular}{cccc}
\includegraphics[width = 0.3 \textwidth]{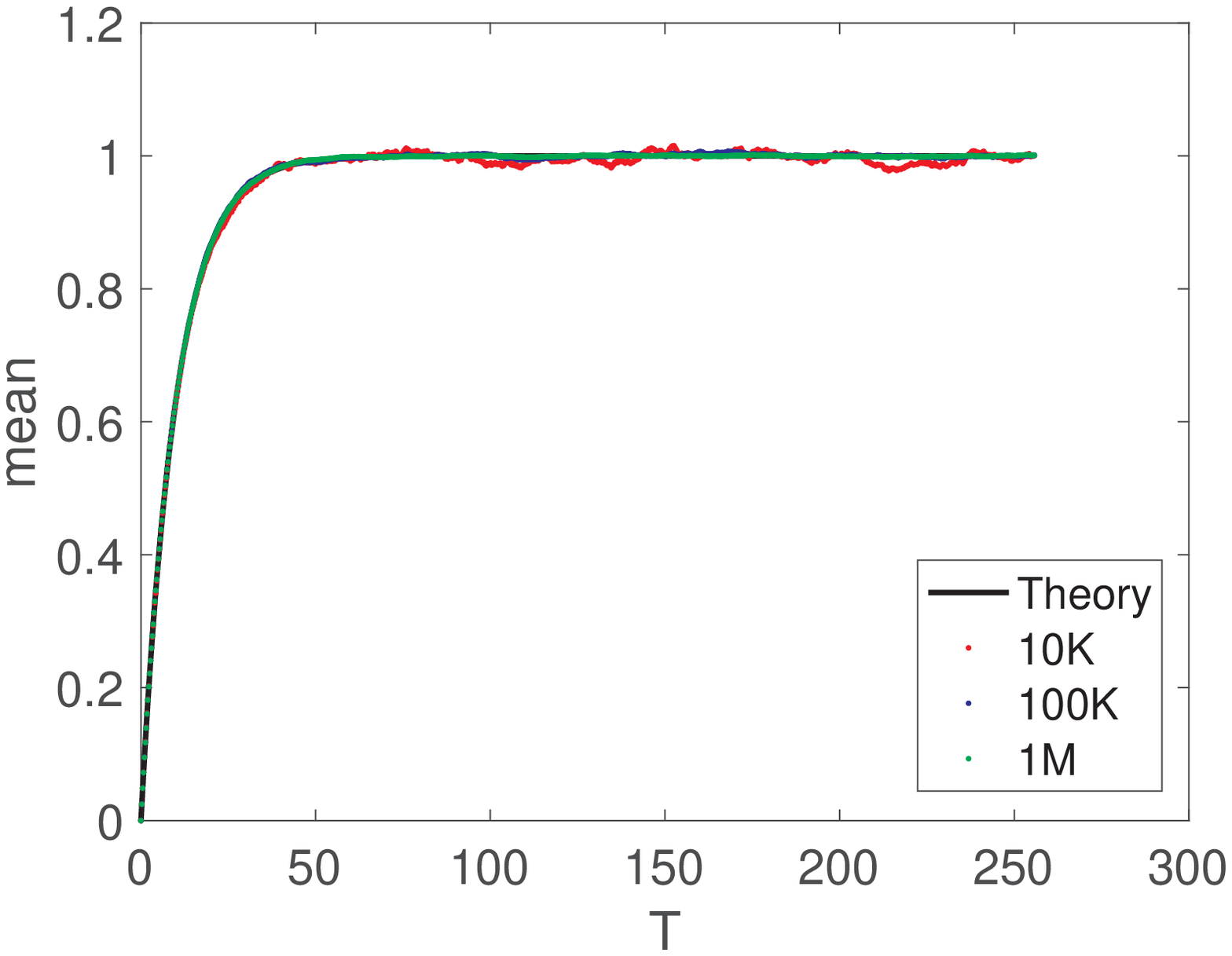}
\includegraphics[width = 0.3 \textwidth]{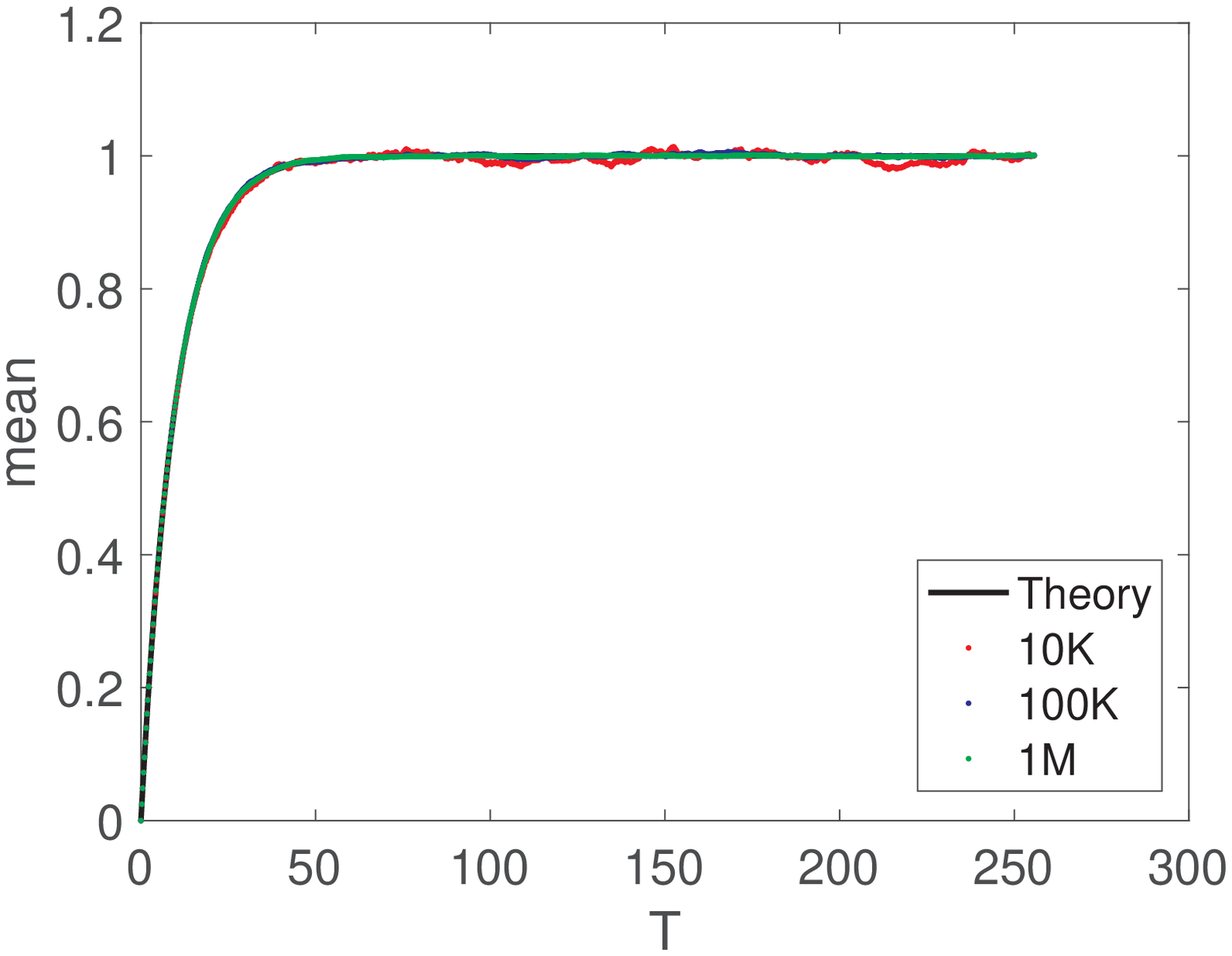}
\includegraphics[width = 0.3 \textwidth]{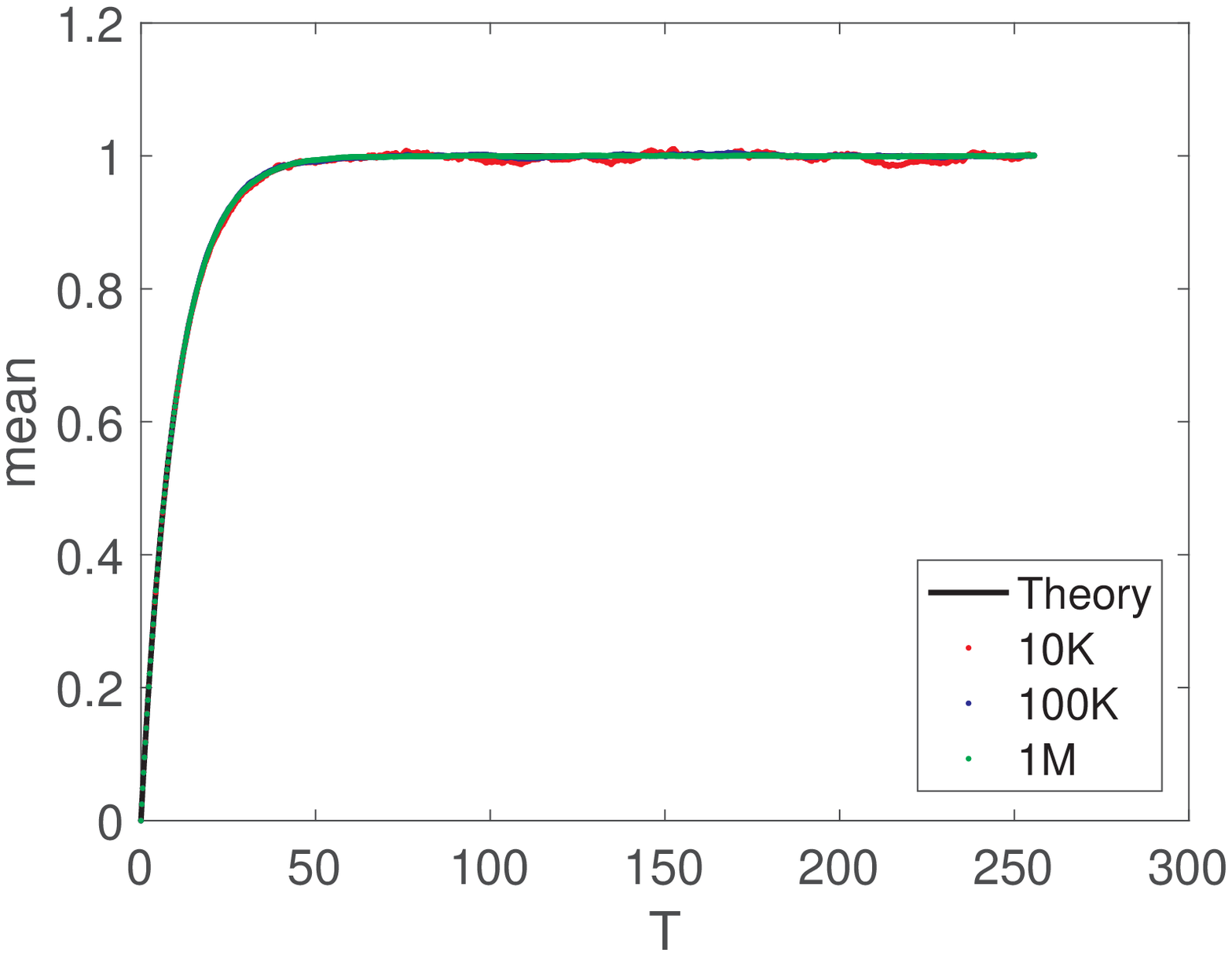}
\end{tabular}
\caption{Saturation of the mean for $x(0)=0$ using time series with duration of $10^4$, $10^5$, and $10^6$ steps vis-a-vis (\ref{x00}); $\gamma=10^{-1}$ and, from left to right, $ \kappa^{2}=10^{-2}$, $\kappa^{2}=8\times10^{-3}$, and $\kappa^{2}=5\times10^{-3}$.}
\label{Meanx0}
\end{figure}

\begin{figure}[!htbp]
\centering
\begin{tabular}{ccc}
\includegraphics[width = 0.3 \textwidth]{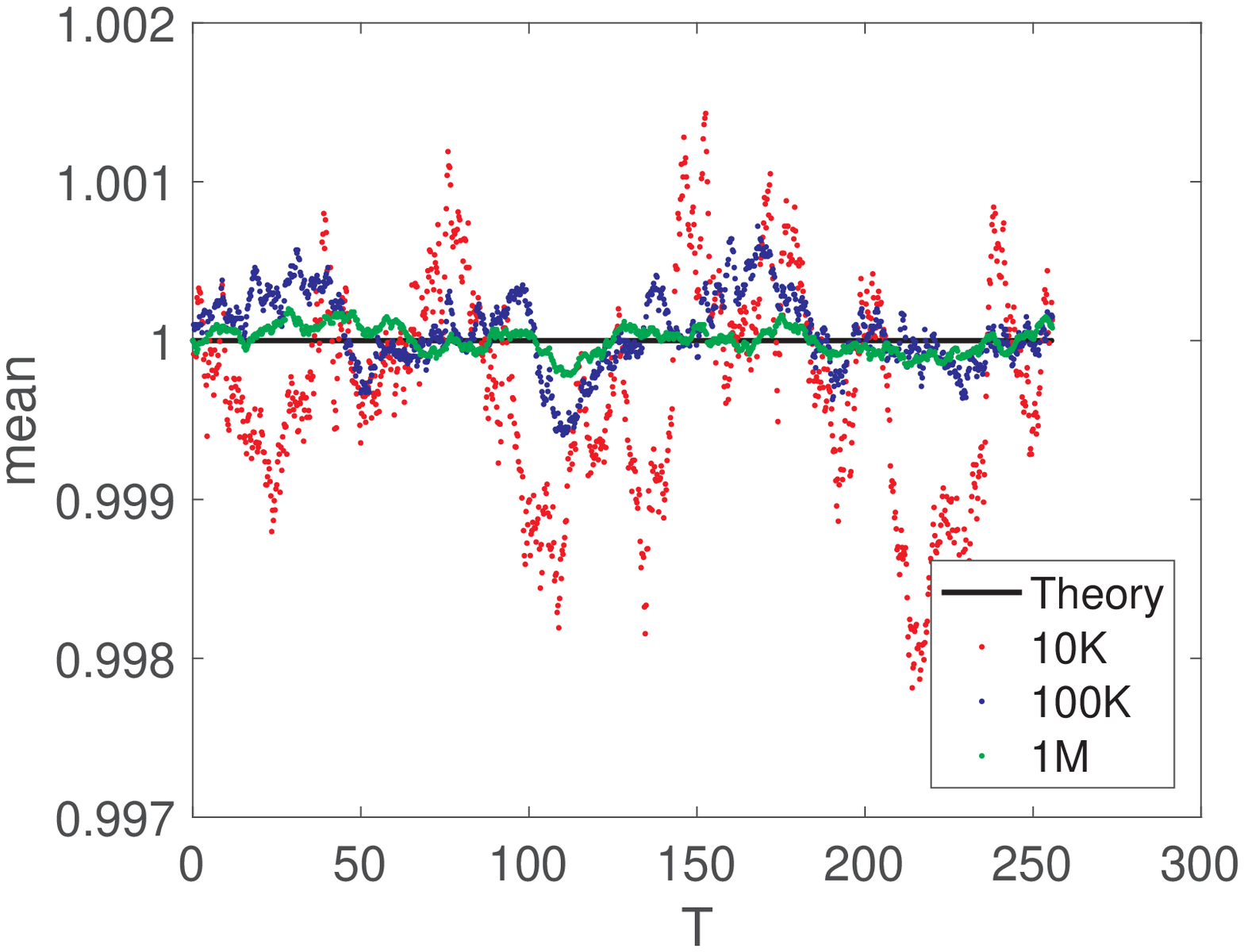}
\includegraphics[width = 0.3 \textwidth]{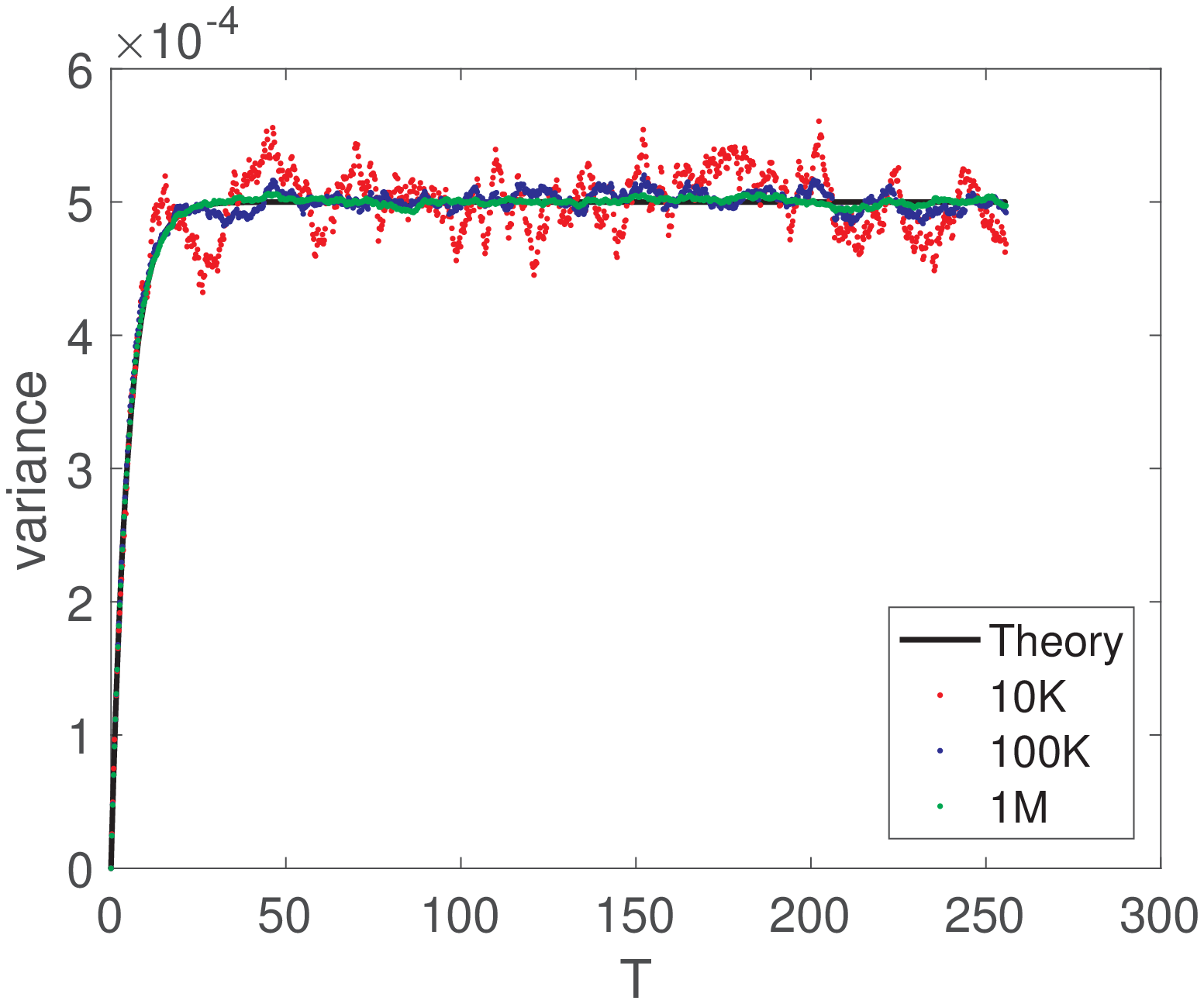}
\includegraphics[width = 0.3 \textwidth]{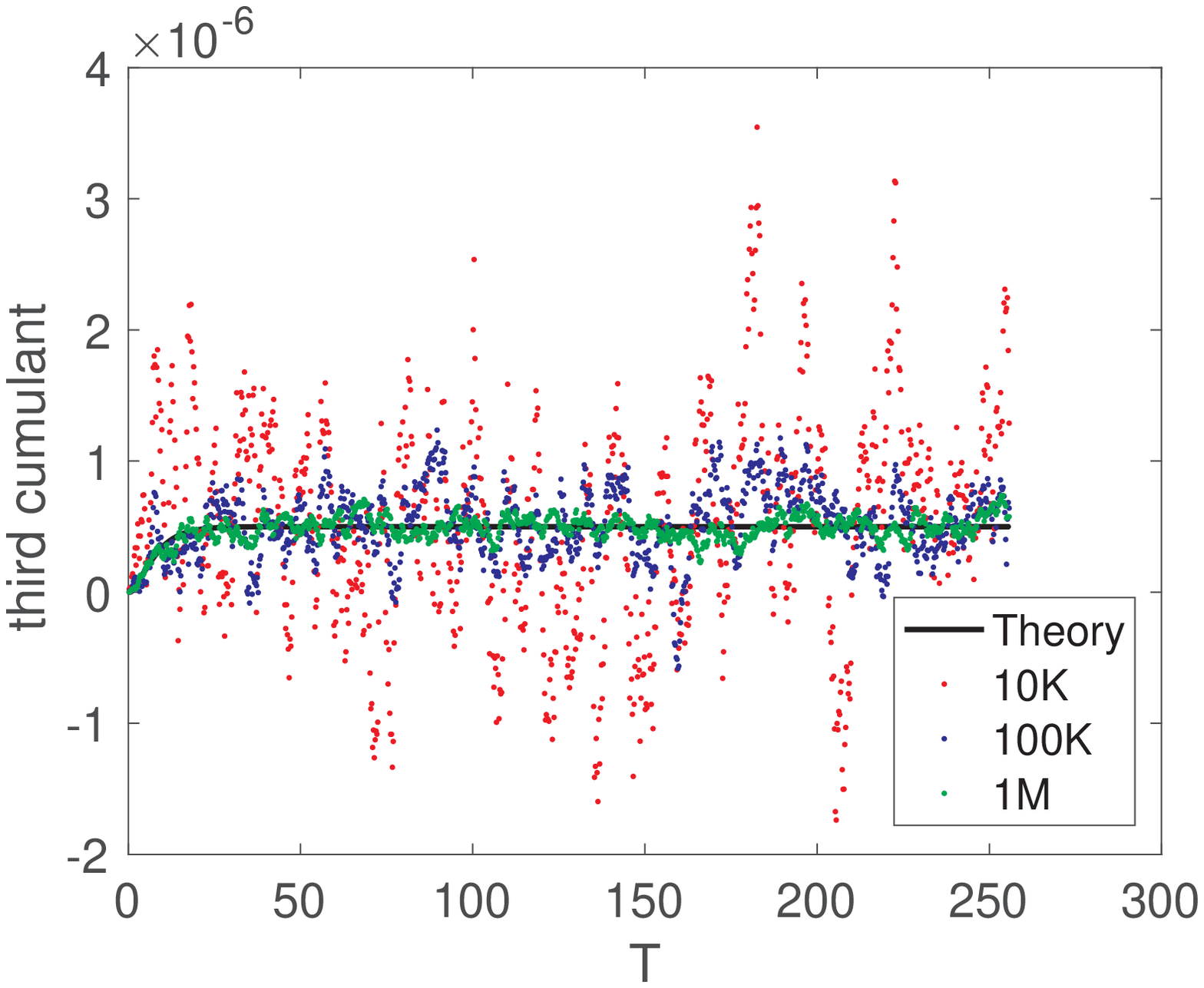}
\end{tabular}
\begin{tabular}{ccc}
\includegraphics[width = 0.3 \textwidth]{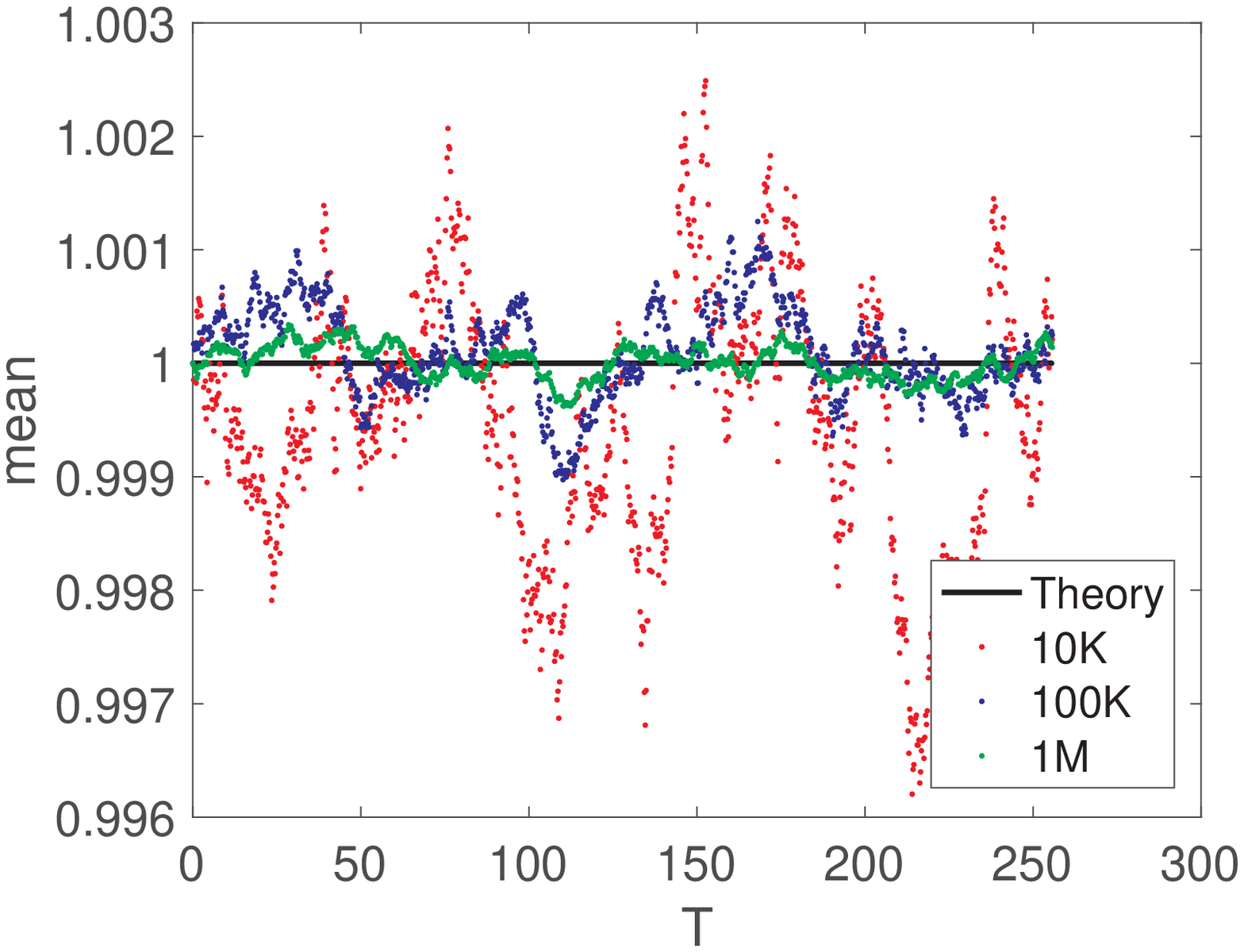}
\includegraphics[width = 0.3 \textwidth]{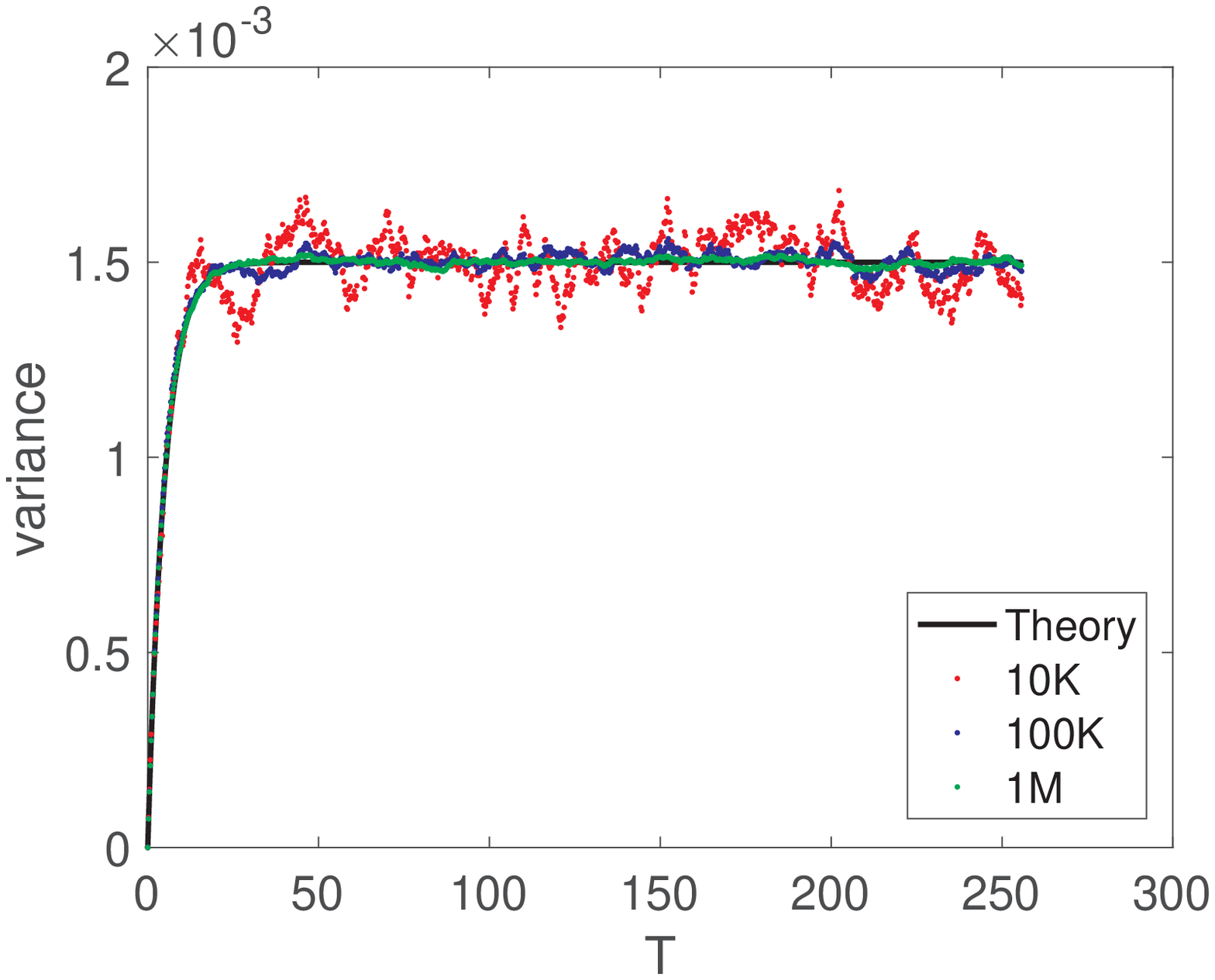}
\includegraphics[width = 0.3 \textwidth]{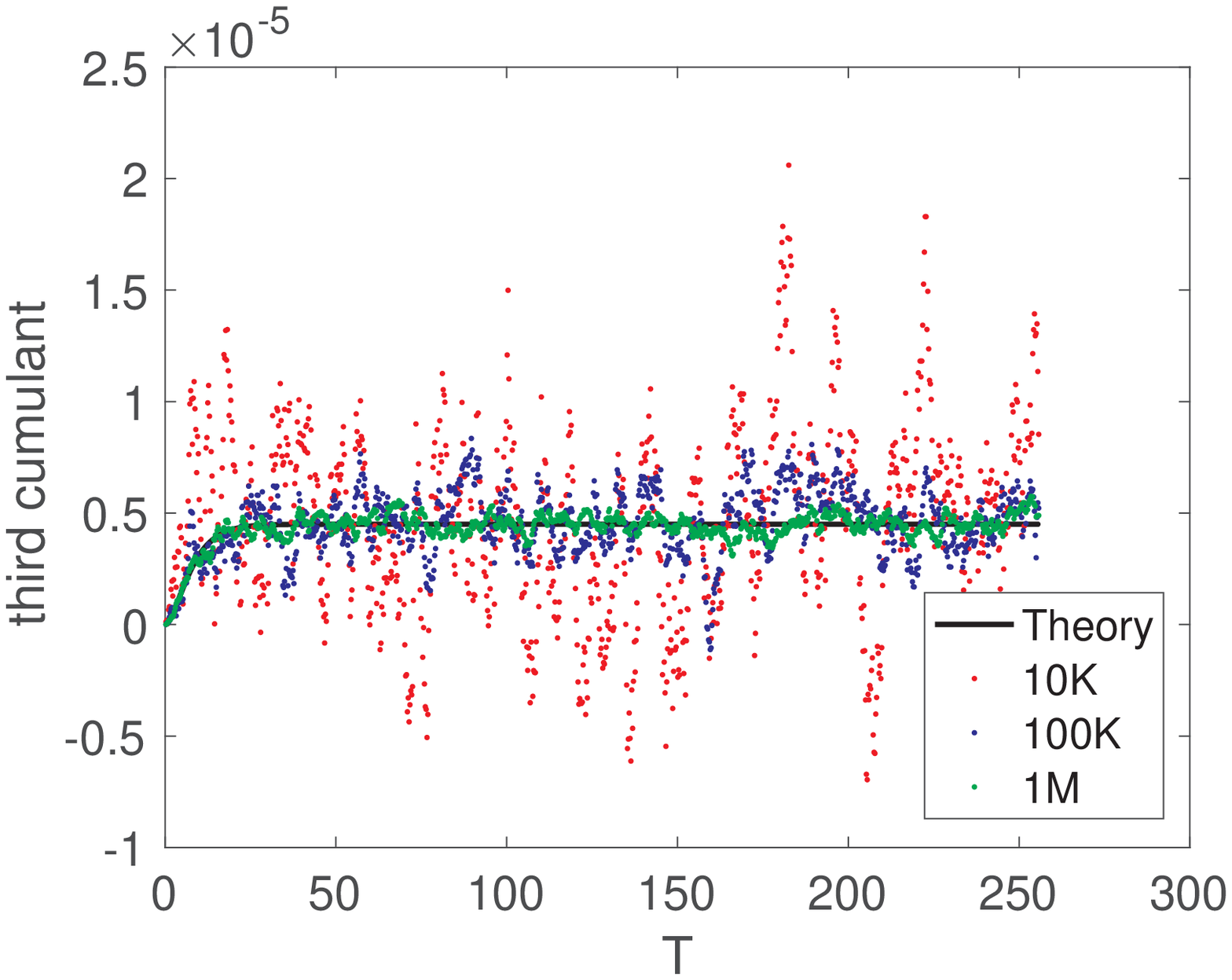}
\end{tabular}
\begin{tabular}{ccc}
\includegraphics[width = 0.3 \textwidth]{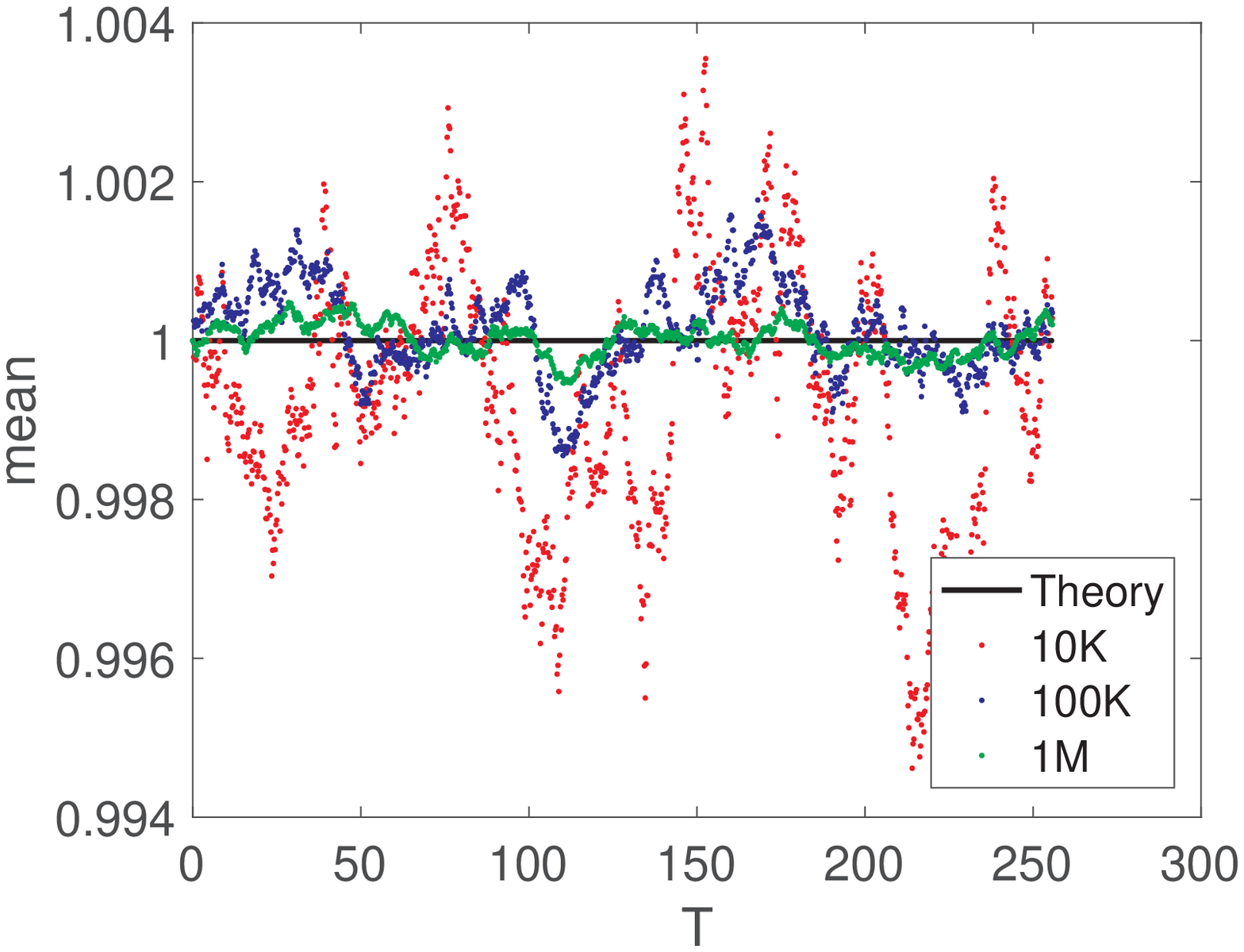}
\includegraphics[width = 0.3 \textwidth]{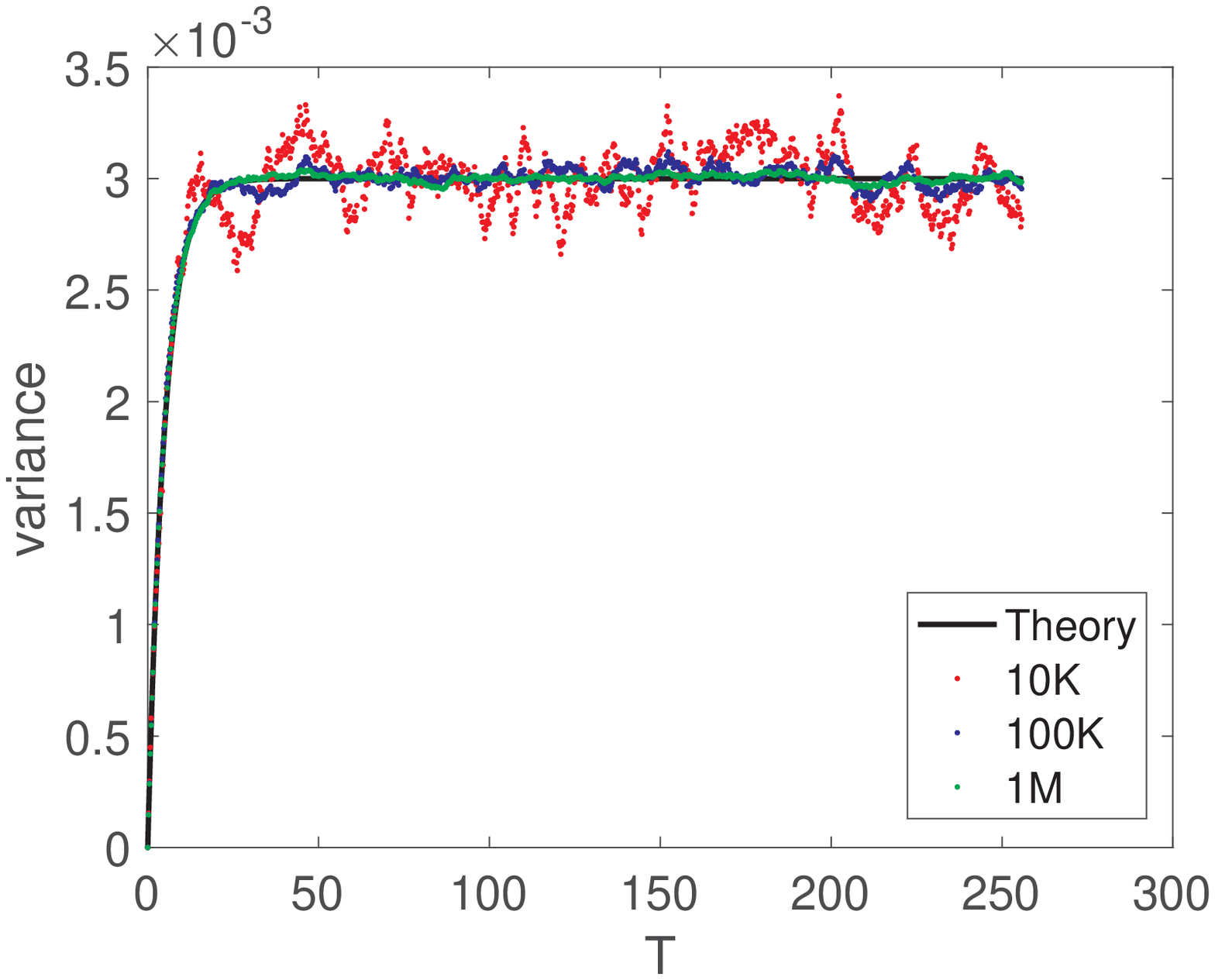}
\includegraphics[width = 0.3 \textwidth]{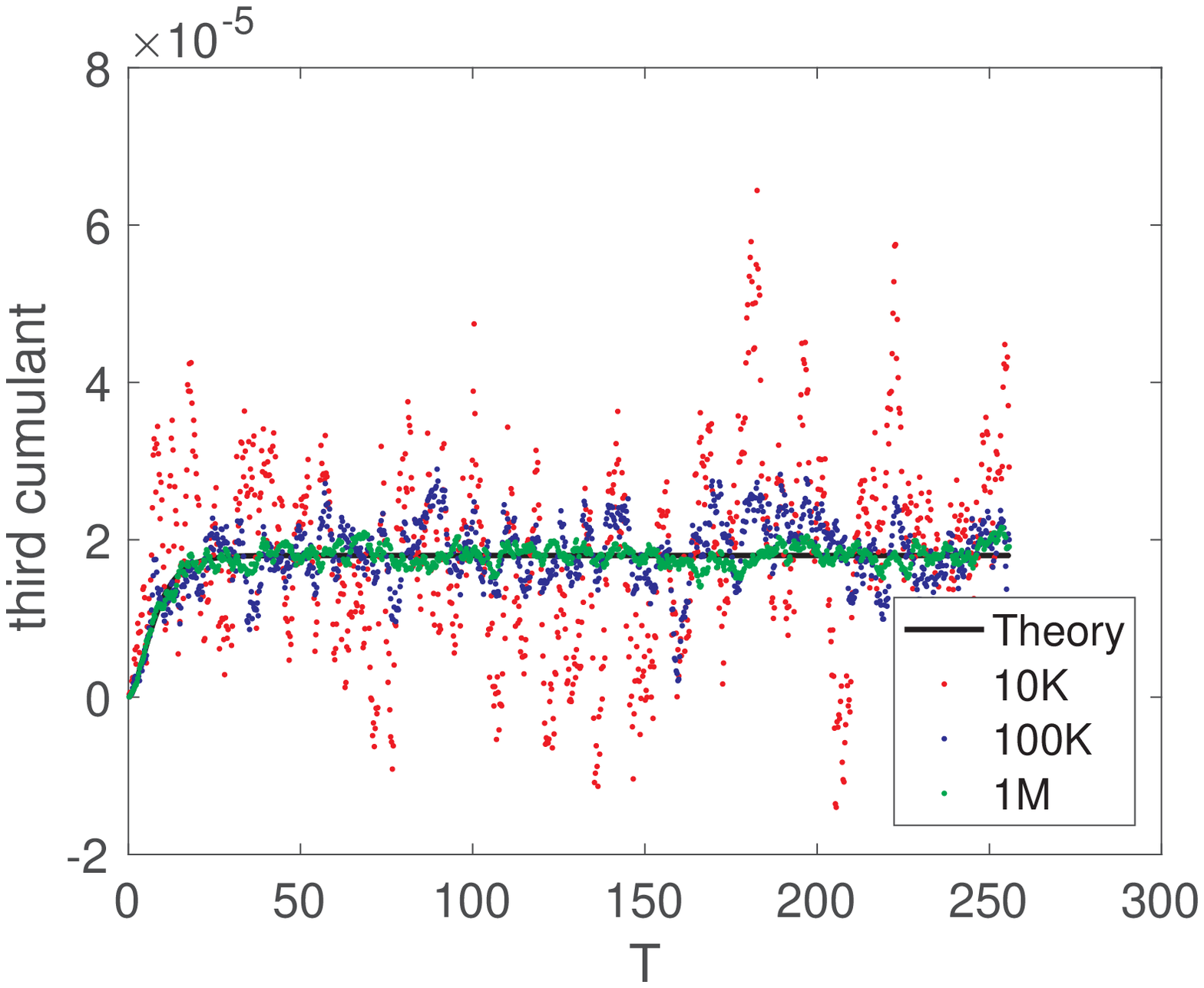}
\end{tabular}
\begin{tabular}{ccc}
\includegraphics[width = 0.3 \textwidth]{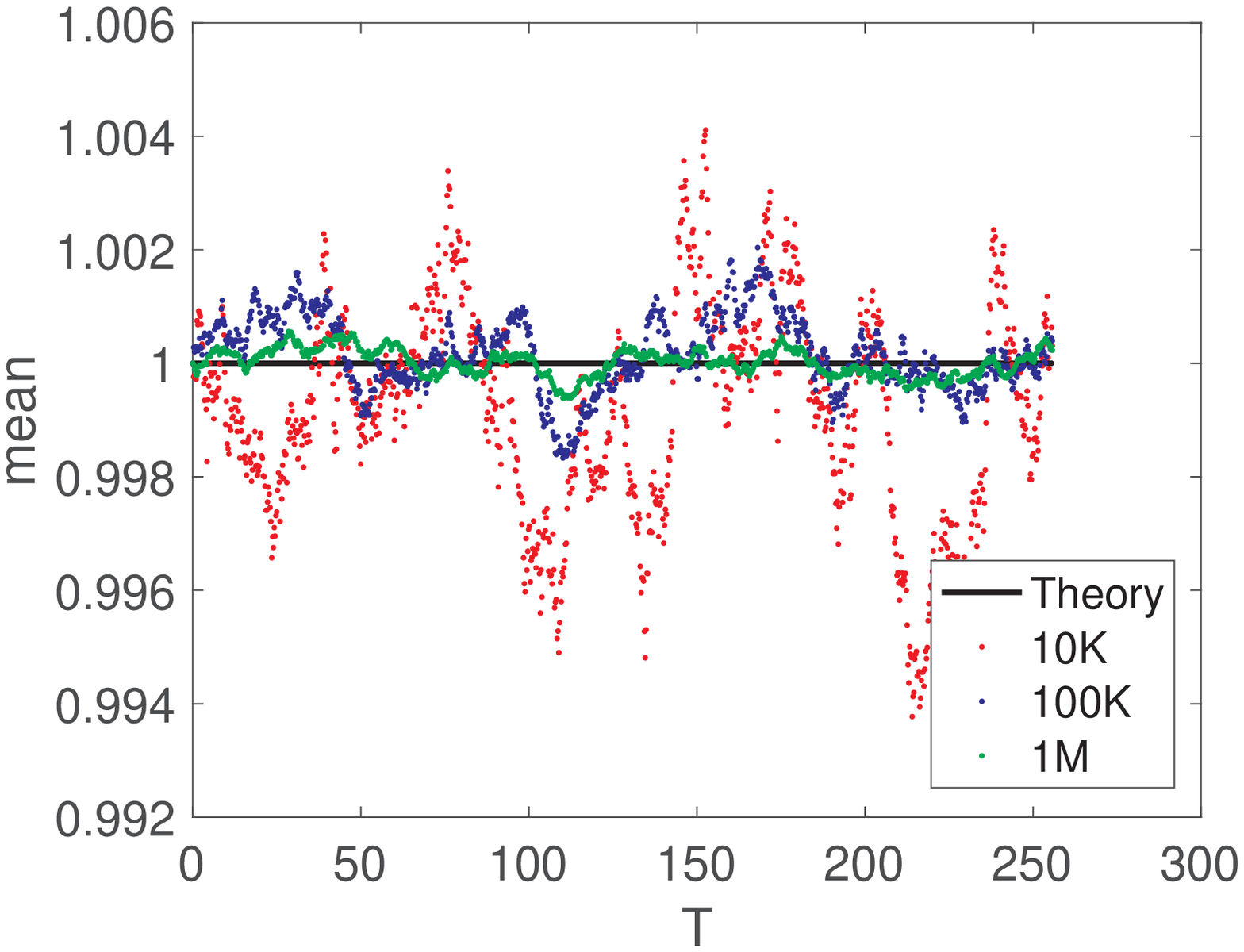}
\includegraphics[width = 0.3 \textwidth]{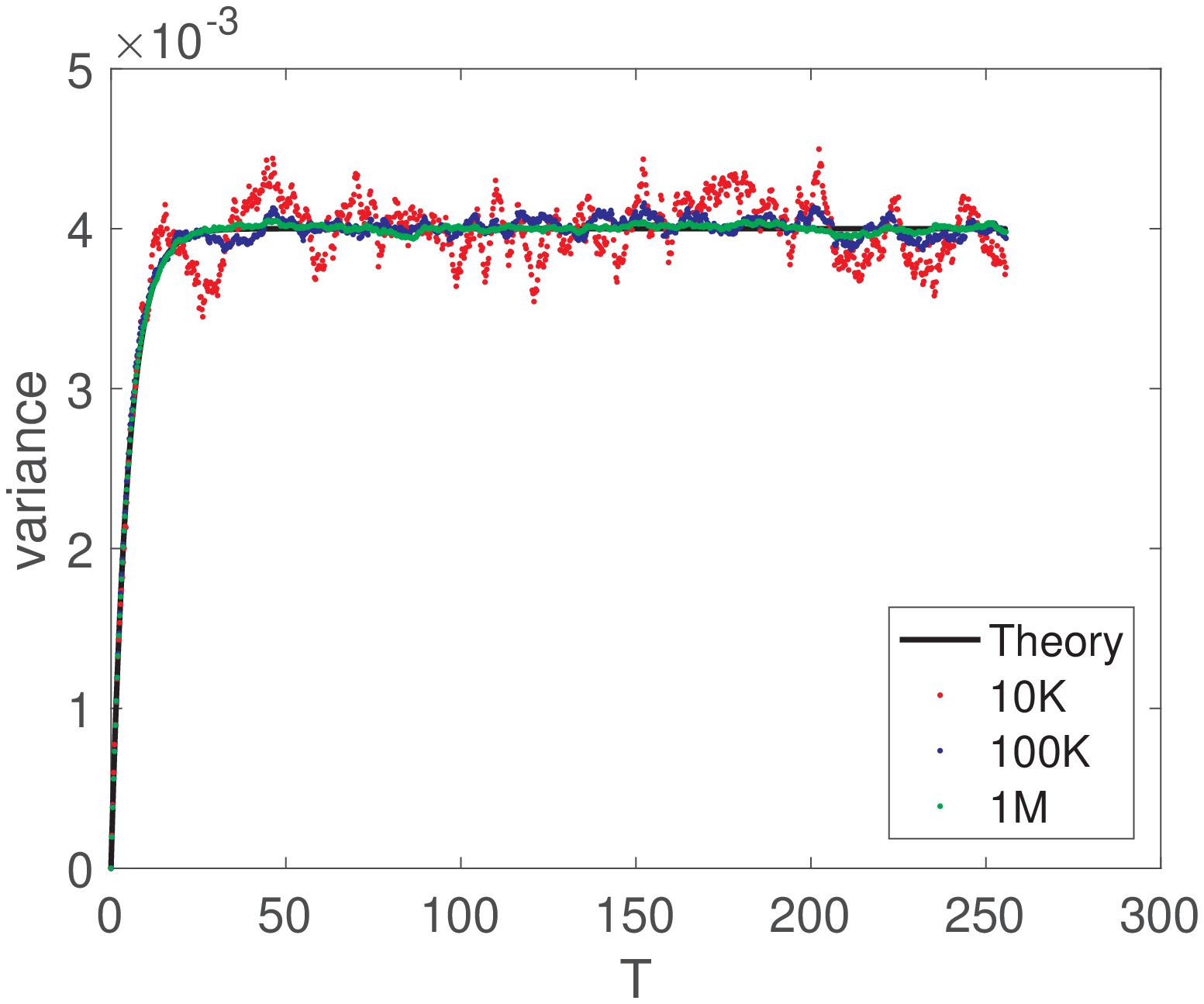}
\includegraphics[width = 0.3 \textwidth]{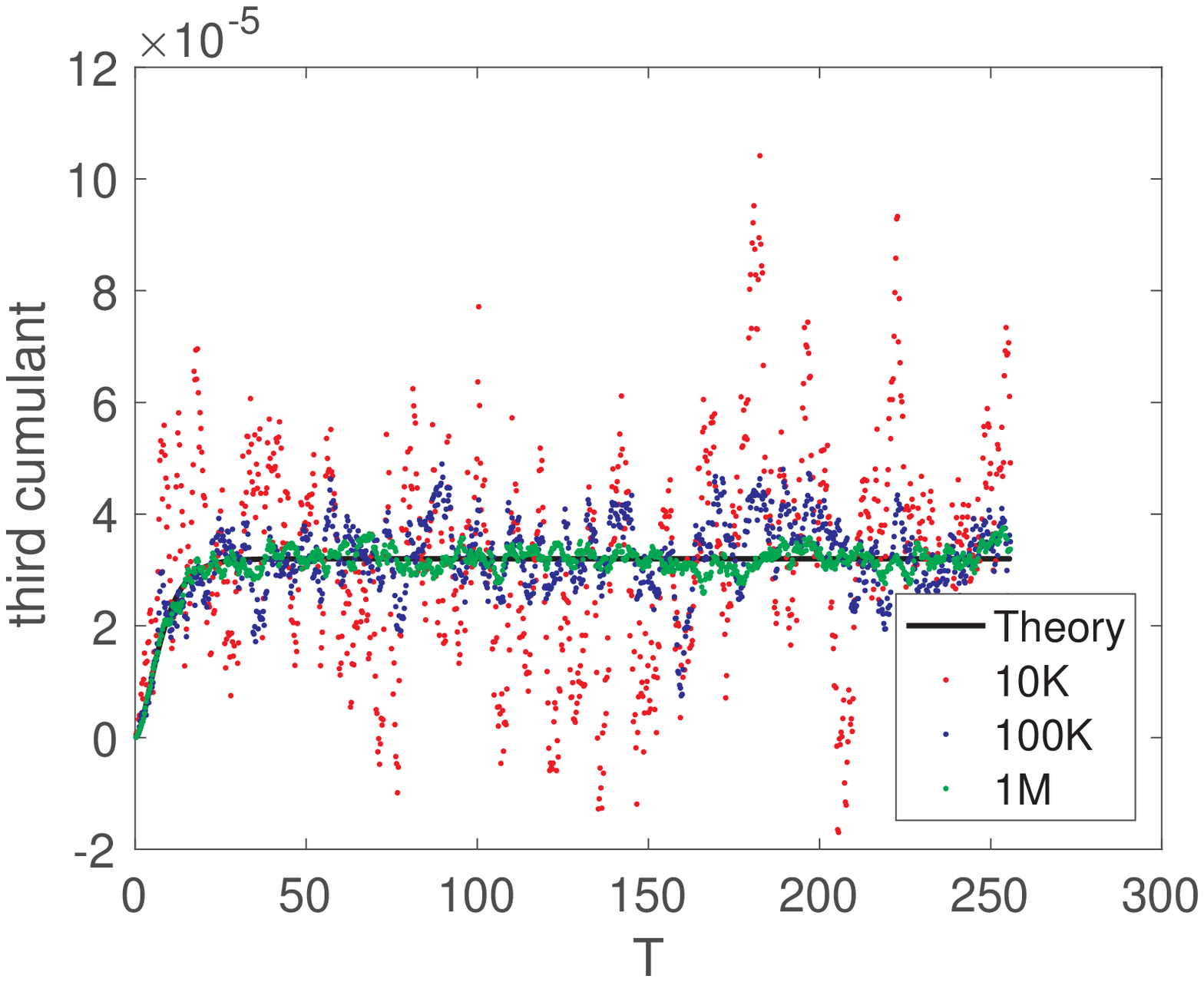}
\end{tabular}
\caption{Saturation of the mean and cumulants for $x(0)=1$ using time series with duration of $10^4$, $10^5$, and $10^6$ steps vis-a-vis (\ref{x01}) and (\ref{x01cumul}); $\gamma=10^{-1}$, and, from top to bottom, $ \kappa^{2}=10^{-4}$, $\kappa^{2}=3\times10^{-4}$, $\kappa^{2}=6\times10^{-4}$, $\kappa^{2}=8\times10^{-4}$.}
\label{MVC1}
\end{figure}

As is for multiplicative model, the easiest way to observe multiple relaxation times predicted using eigenvalue method, is through relaxation of cumulants \cite{liu2017correlation}. As was observed in \ref{FPapproach}, the mean can always be set to unity, $\theta=1$, and in what follows we will use (\ref{CIR1}). We will also use two sets of initial conditions,  $x(0)=0$ and $x(0)=1$. For the former, the expressions for the mean and the cumulants are given by
\begin{equation}
x(0)=0: \hspace{0.21cm} <x>=1-e^{-\gamma t}, \hspace{0.21cm}
\kappa_n=(\frac{\kappa^2}{\gamma})^{n-1} \frac{(n-1)!}{2^{n-1}} e^{-n\gamma t}(e^{\gamma t} - 1)^n
\label{x00}
\end{equation}
and in particular
\begin{equation}
x(0)=0: \hspace{0.21cm} \kappa_2=\frac{\kappa^2}{2\gamma}(1-2 e^{-\gamma t}+e^{-2\gamma t}), \hspace{0.21cm}
\kappa_3=\frac{\kappa^4}{2\gamma^2}(1-3 e^{-\gamma t}+3 e^{-2\gamma t}- e^{-3\gamma t})
\label{x00cumul}
\end{equation}
For the latter, we have
\begin{equation}
x(0)=1: \hspace{0.21cm} <x>=1,  \hspace{0.21cm}
\kappa_n=(\frac{\kappa^2}{\gamma})^{n-1} \frac{(n-1)!}{2^{n-1}} e^{-n\gamma t}(e^{\gamma t} - 1)^{n-1} (e^{\gamma t} +(n-1))
\label{x01}
\end{equation}
and in particular
\begin{equation}
x(0)=1: \hspace{0.21cm} \kappa_2=\frac{\kappa^2}{2\gamma}(1 - e^{-2t\gamma}), \hspace{0.21cm}
\kappa_3=\frac{\kappa^4}{2\gamma^2} (1 - 3 e^{-2\gamma t} + 2 e^{-3\gamma t})
\label{x01cumul}
\end{equation}
The behavior of the mean starting with $x(0)=0$ is shown in Fig. \ref{Meanx0} and the behavior of the mean and cumulants $\kappa_2$ and $\kappa_3$ starting with $x(0)=1$ in Fig. \ref{MVC1}. Time series of various durations were used, as well as different values of $\kappa^2$. Clearly, theory describes the mean and cumulants approach to equilibrium values very well.  

\subsection{Distribution of Relaxation Times\label{DistRelaxTimes}}
\begin{figure}[!htbp]
\centering
\begin{tabular}{cc}
\includegraphics[width = 0.61 \textwidth]{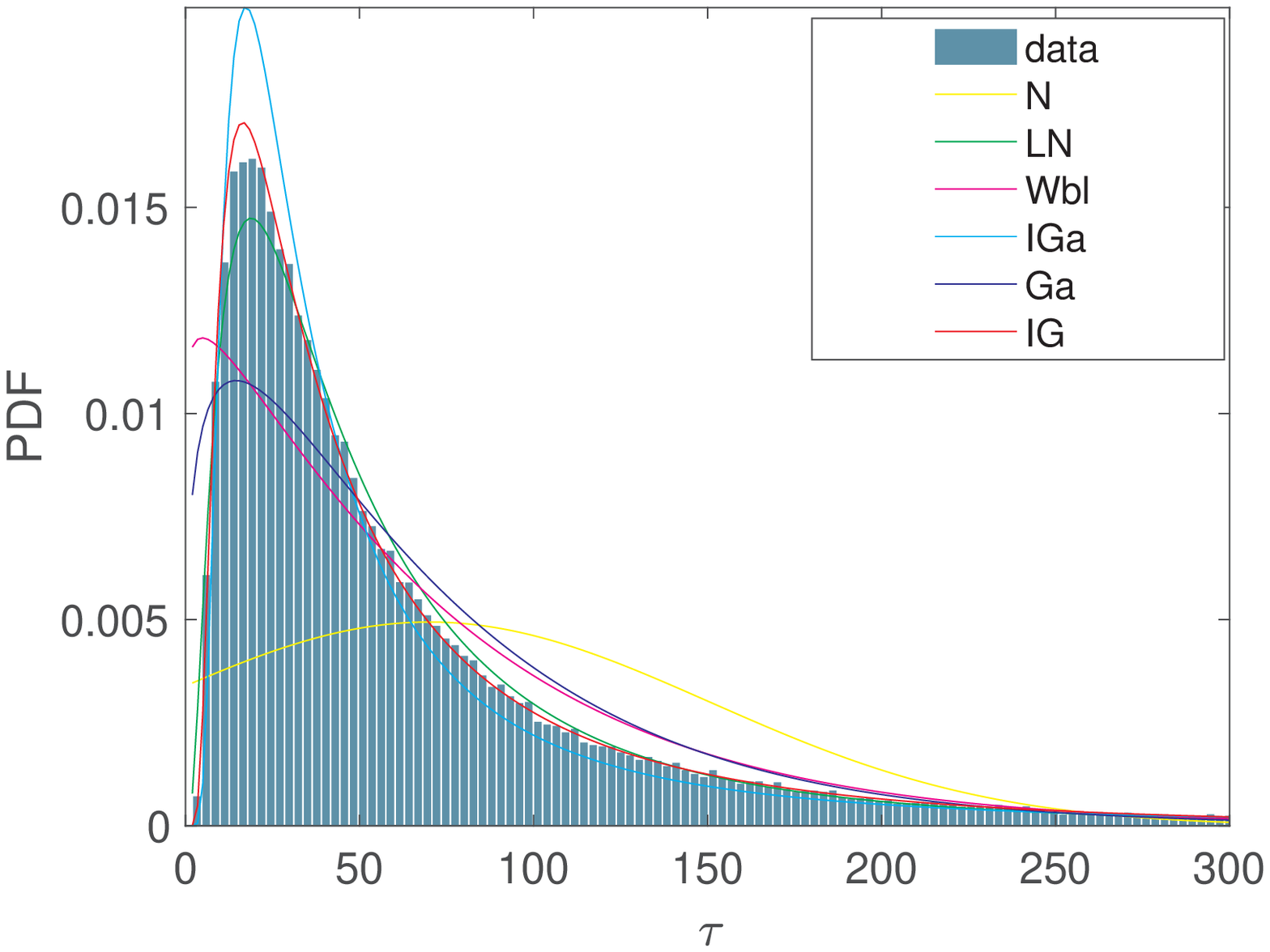}
\end{tabular}
\begin{tabular}{cc}
\includegraphics[width = 0.61 \textwidth]{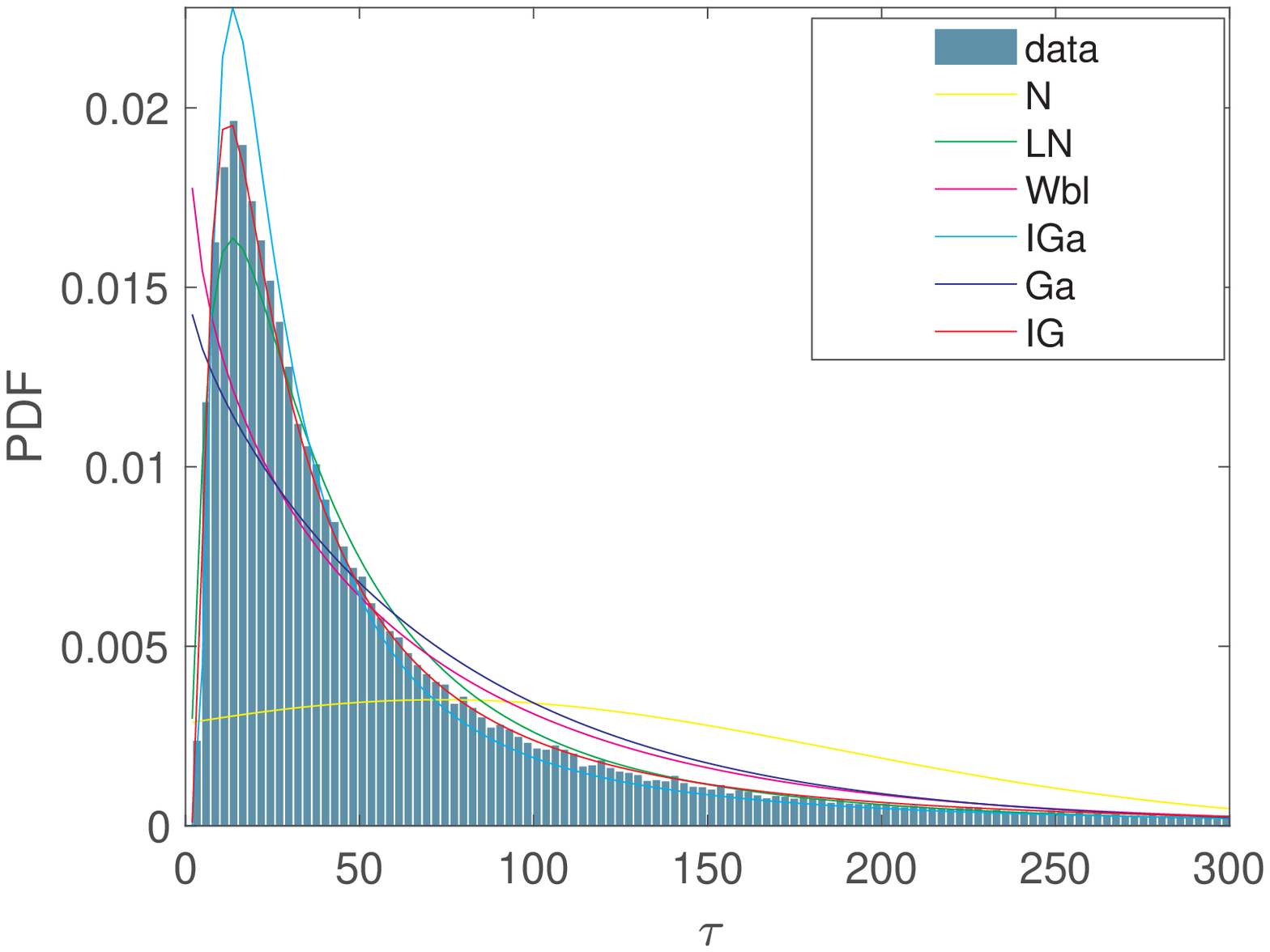}
\end{tabular}
\caption{Fits of the distribution of relaxation times from Table \ref{TableDRT} for $\kappa^{2}=10^{-4}$ (top) and $\kappa^{2}=0.5\times10^{-1}$ (bottom)}
\label{fitsDRT}
\end{figure}

In the same manner as was done for multiplicative model \cite{liu2017correlation}, we investigate the distribution of relaxations times. Namely, we generate a time series (\ref{CIR1}) and observe how quickly its distribution approaches the steady-state distribution (\ref{SSrelax}). The relaxation time is determined by saturation of the Kolmogorov-Smirnov (KS) statistic for comparison between numerical and theoretical distribution to its lowest value. We generated $10^5$ relaxation times and studies their distribution function. We fitted with Normal (N), Lognormal (LN), InverseGamma (IGa), Gamma (Ga), Weibull (Wbl) and Inverse Gaussian (IG) distributions using maximum likelihood estimation (MLE) and evaluated KS statistics for this fits (lower KS numbers indicate better fits.) The results are summarized in Table \ref{TableDRT} and fits, for the same $\gamma$ as in Table \ref{TableDRT} and two values of $\kappa^2$ from it, are shown in Fig. \ref{fitsDRT}.

\begin{table}[!htbp]
\centering
\caption{MLE-obtained parameters and KS values for fitting distribution function of relaxation times for $\gamma=10^{-1}$ and several values of $\kappa^2$.}
\label{MLEand KS}
\begin{tabular}{cccc}
\hline
			$\kappa^{2}=10^{-4}$&      &   $\kappa^{2}=10^{-3}$&    \\
\hline
            parameters &          KS test&		parameters &          KS test\\
\hline
 N(70.0421,80.7143) &           0.2071 & N(70.4884,86.8047) \\
\hline
 LN(          3.8031,           0.9317) &           0.0175 & LN(          3.7814,           0.9532) &           0.0198   \\
\hline
 IGa(          1.3787,          41.2501) &           0.0455 & IGa(          1.3383,          38.6621) &           0.0437  \\
\hline
 Gamma(          1.2618,          55.5088) &           0.0806 & Gamma(          1.1940,          59.0379) &           0.0859  \\
\hline
 Weibul(         71.9396,           1.0601) &           0.0703 & Weibul(         71.2892,           1.0235) &           0.0746 \\
\hline
 IG(         70.0421,          52.2334) &           0.0083 & IG(         70.4884,          48.9500) &           0.0066  \\
\hline
\end{tabular}
\begin{tabular}{cccc}
\hline
			$\kappa^{2}=2\times10^{-2}$ &    & $\kappa^{2}=1.5\times10^{-2}$ &\\
\hline
            parameters &          KS test&		parameters & KS test\\
\hline
  N(71.0994,93.2400) &           0.2350 & N(70.7593, 89.8287) &           0.2281 \\
\hline
 LN(          3.7596,           0.9759) &           0.0226 &  LN(          3.7699,           0.9652) &           0.0211 \\
\hline
IGa(          1.2983,          36.1767) &           0.0418 &  IGa(          1.3162,          37.2958) &           0.0431\\
\hline
 Gamma(          1.1286,          62.9993) &           0.0922 & Gamma(          1.1599,          61.0050) &           0.0889 \\
\hline
 Weibul(         70.6876,           0.9888) &           0.0783 & Weibul(         70.9619,           1.0057) &           0.0764  \\
\hline
  IG(         71.0994,          45.8257) &           0.0088 & IG(         70.7593,          47.2629) &           0.0070 \\
\hline
\end{tabular}
\begin{tabular}{cccccc}
\hline
   $\kappa^{2}=0.5\times10^{-1}$ &        &   $\kappa^{2}=5.6\times10^{-2}$ &\\
\hline
parameters &          KS test&			parameters &          KS test\\
\hline
 N(73.3633,113.5198) &           0.2678 & N(73.8387,117.0505) &           0.2723\\
\hline
  LN(          3.6953,           1.0479) &           0.0291 &  LN(          3.6839,           1.0612) &           0.0304 \\
\hline
 IGa(          1.1737,          29.1487) &           0.0387 & IGa(          1.1534,          28.0570) &           0.0381 \\
\hline
Gamma(          0.9659,          75.9566) &           0.1058 & Gamma(          0.9409,          78.4754) &           0.1078 \\
\hline
 Weibul(         69.0188,           0.9031) &           0.0863 & Weibul(         68.7421,           0.8900) &           0.0877  \\
\hline
 IG(         73.3633,          37.5457) &           0.0136  & IG(         73.8387,          36.2745) &           0.0142 \\
\hline
\end{tabular}
\label{TableDRT}
\end{table}

\begin{figure}[!htbp]
\centering
\begin{tabular}{ccc}
\includegraphics[width = 0.3 \textwidth]{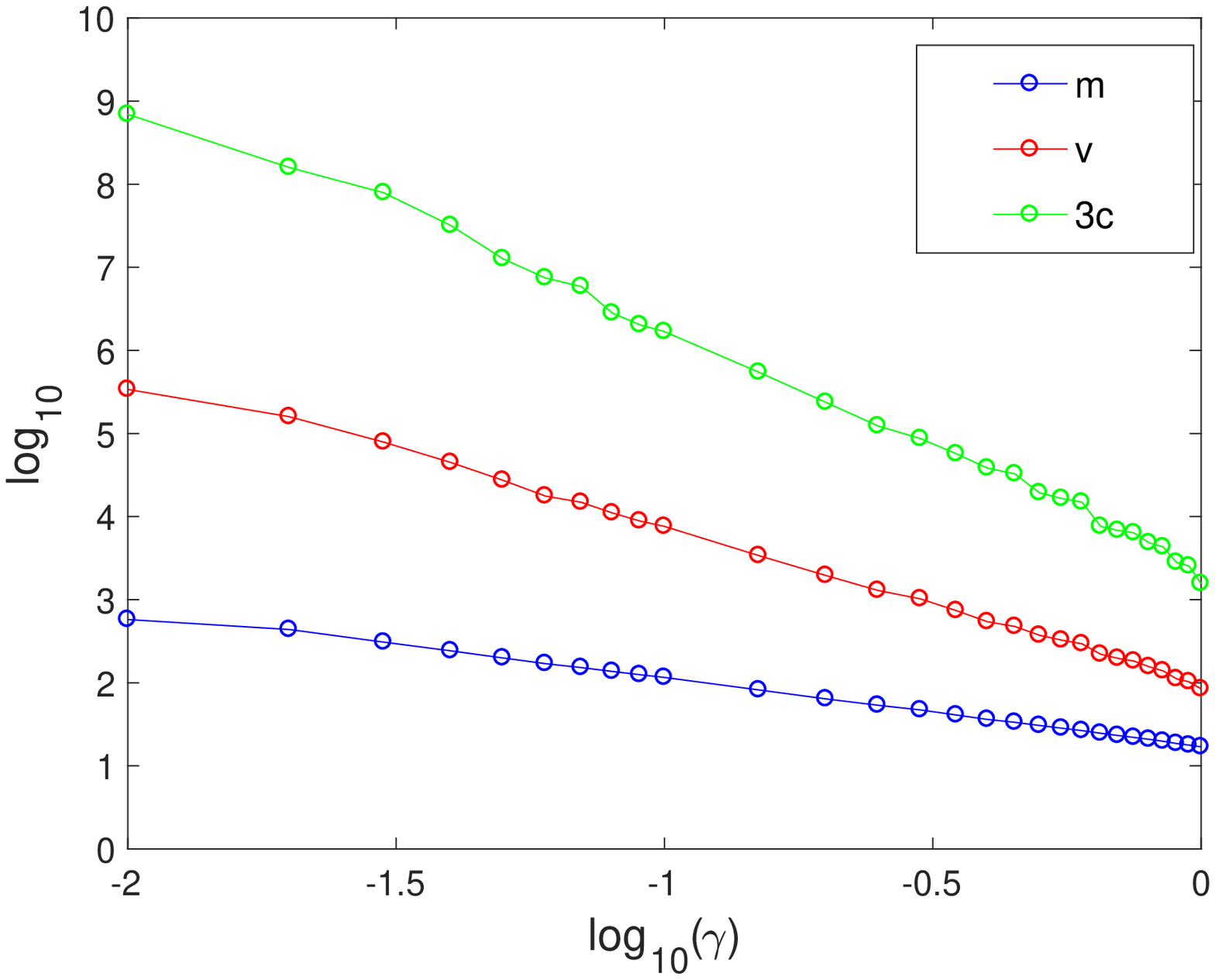}
\includegraphics[width = 0.3 \textwidth]{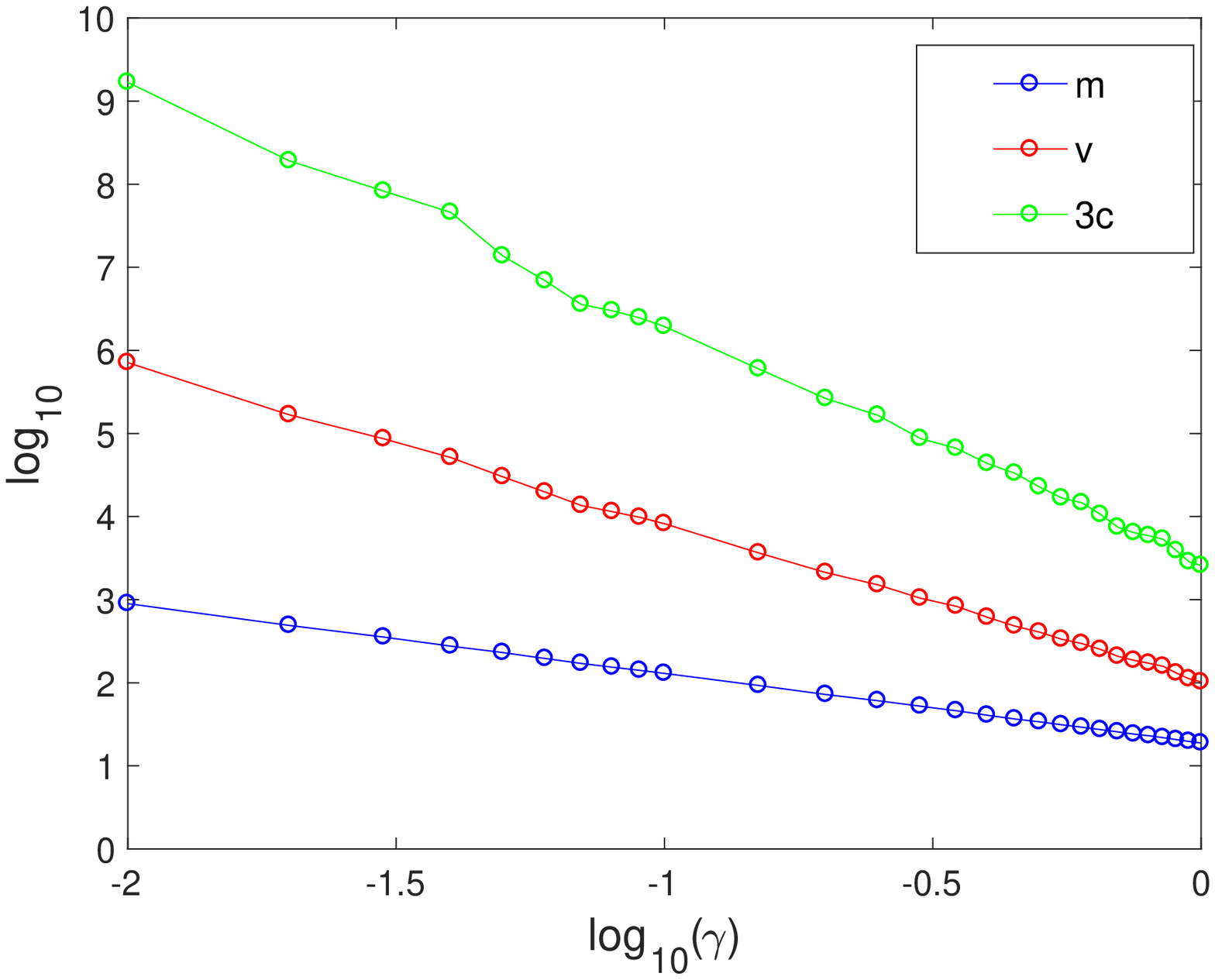}
\includegraphics[width = 0.3 \textwidth]{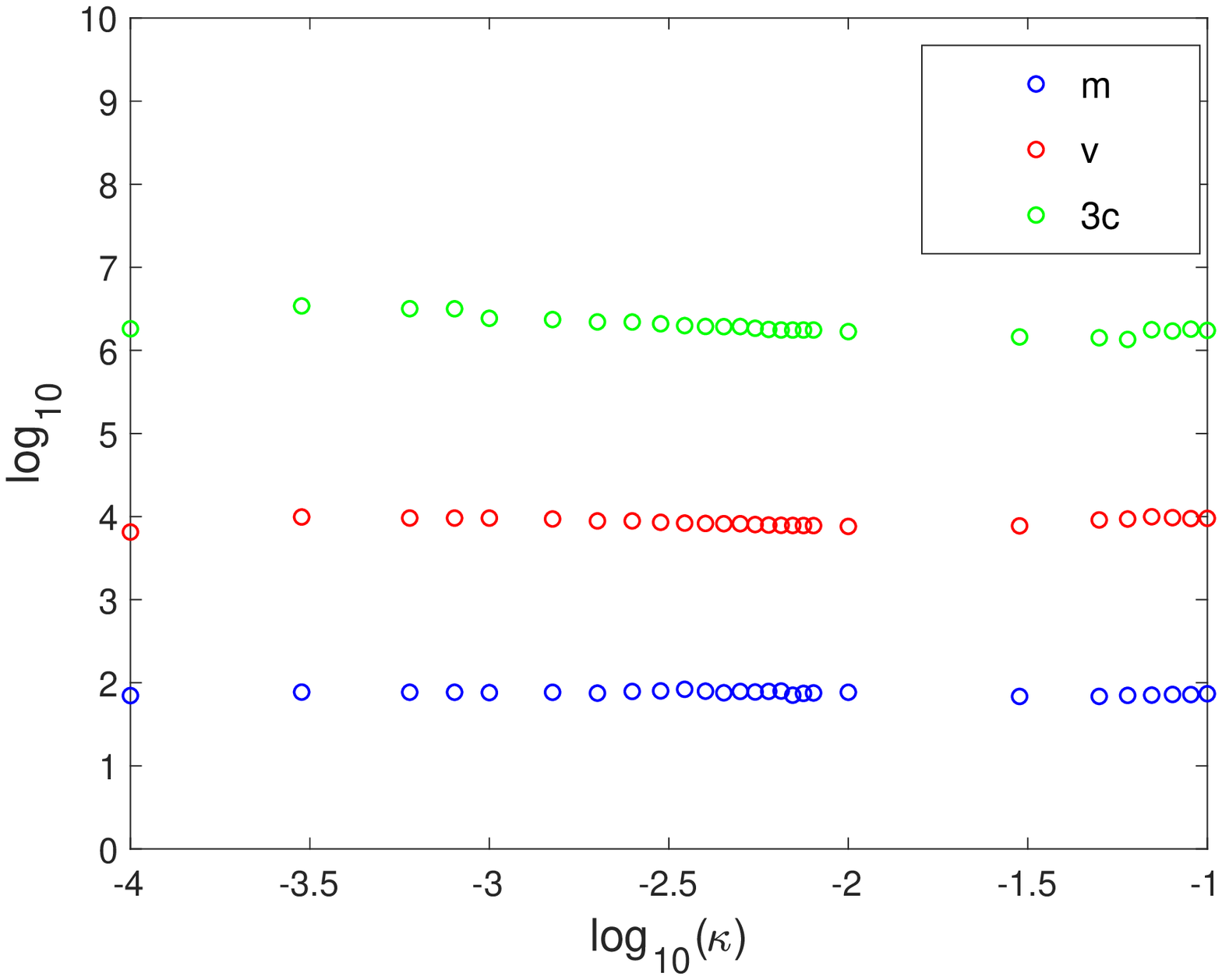}
\end{tabular}
\caption{On log-log scale, dependence of the mean, variance and third cumulant of the relaxation time distribution on $\gamma$ for $\kappa^2=10^{-2}$ (left) and $\kappa^2=5 \times 10^{-3}$ (middle) for $\gamma$ varying between $10^{-2}$ and $1$. For $\gamma=10^{-1}$ dependence on $\kappa$ which varies between $10^{-4}$ and $0.1$ (right).}
\label{cumulIG}
\end{figure}

As was the case with the multiplicative model, IG distribution
\begin{equation}
IG(a\gamma^{-1},b\gamma^{-1};x)=\sqrt{\frac{b}{2\pi \gamma x^3}} exp[-\frac{b\gamma (x-a\gamma^{-1})^2}{2a^2x}]
\label{IGDist}
\end{equation}
provided by far the best fit. It should be noted that for (\ref{IGDist}), cumulants $\kappa_n \propto \gamma^{-n}$ and are independent of the coefficient $\kappa$. Fig. \ref{cumulIG} shows excellent agreement with numerical results.


\clearpage

\bibliography{mybib}

\end{document}